\documentclass[11pt]{article}
\usepackage[a4paper,margin=0.75in]{geometry}
\usepackage{amsmath,amssymb,amsthm}
\usepackage{graphicx,booktabs,array}
\usepackage[round,sort]{natbib}
\usepackage{hyperref}
\usepackage{bm}
\usepackage{enumerate}
\usepackage{setspace}
\usepackage{subcaption}
\newtheorem{theorem}{Theorem}
\newtheorem{lemma}[theorem]{Lemma}
\newtheorem{proposition}[theorem]{Proposition}

\newtheorem{remark}{Remark}
\theoremstyle{definition}
\newtheorem{definition}{Definition}
\newtheorem{example}{Example}

\newcommand{\bpsi}{\boldsymbol{\psi}}
\newcommand{\btheta}{\boldsymbol{\theta}}
\newcommand{\bZ}{\boldsymbol{Z}}
\newcommand{\bz}{\boldsymbol{z}}
\newcommand{\bV}{\boldsymbol{V}}
\newcommand{\bY}{\boldsymbol{Y}}
\newcommand{\bX}{\boldsymbol{X}}
\newcommand{\bx}{\boldsymbol{x}}
\newcommand{\bw}{\boldsymbol{w}}
\newcommand{\E}{\mathbb{E}}

\newcommand{\NS}{\mathrm{NS}}

\title{Fractionally Supervised Classification with Maxima Nominated Samples}
\author{\normalsize
Mohammad Jafari Jozani $^{a,}$\footnote{Corresponding author. Email:  \texttt{m\_jafari\_jozani@umanitoba.ca}}~~ and Jingyu Wang $^b$\\
\footnotesize{$^a$ Department of Statistics,  University of Manitoba,
                        Winnipeg, MB, Canada} \\
\footnotesize{$^b$ RBC Capital Markets, Toronto, Canada }}

\date{\today}

\begin{document}
\maketitle

%=========================================================================
\begin{abstract}
\noindent

Fractionally supervised classification (FSC) offers a flexible framework for combining labeled and unlabeled data in model-based classification, but existing formulations assume simple random sampling. In many applications, however, the retained observation is an extreme order statistic from a set rather than a randomly selected unit. This is particularly appealing when the target population is rare, since maxima nomination sampling (NS) can enrich the sample with the most informative observations, as in screening, environmental monitoring, repeated testing, and reliability studies. Under such designs, the likelihood function changes fundamentally, and the usual FSC EM construction is no longer valid. We develop FSC for nominated samples by introducing a latent representation that accounts for both the class membership of the observed maximum and the latent composition of the remaining units in the set. The resulting method yields a proper EM algorithm and a coherent weighted-likelihood FSC procedure for NS data. We present the methodology in general form, illustrate it for a rare-event contamination normal mixtures, and show through simulation that it substantially improves on the misspecified alternative by ignoring the extra rank information of such data. A real-data analysis demonstrates its practical value.

\end{abstract}

%=========================================================================
\section{Introduction}
\label{sec:intro}
%=========================================================================

Classification problems with only partial labeling arise frequently in modern
statistical practice. In many studies, measurements are readily available,
whereas reliable class labels are expensive, delayed, or available only for a
limited subset of units. This imbalance naturally motivates methods that can
exploit unlabeled observations without discarding the stabilizing information
carried by the labeled sample. Within model-based classification, this issue
has traditionally been studied through supervised, semi-supervised, and
unsupervised paradigms, depending on the extent to which unlabeled data are
used in estimation and prediction \citep{bouveyron2019,dean2006}.

A particularly flexible development in this direction is fractionally
supervised classification (FSC), introduced by \citet{vrbik}, which places
supervised, semi-supervised, and unsupervised classification on a common
continuum through a weighted likelihood. Rather than fixing the role of the
unlabeled sample in advance, FSC allows its contribution to be regulated
explicitly, thereby providing a principled way to interpolate between the
classical regimes. This flexibility is especially attractive in partially
labeled mixture problems, where the amount of supervision appropriate for
estimation is rarely clear a priori. Subsequent work has further developed the
framework through weight selection and extensions beyond Gaussian mixtures; see
\citet{gallaugher2019}.

Although FSC provides a flexible way to combine labeled and unlabeled data,
its current formulation is rooted in simple random sampling. That assumption
is often implicit, but it is fundamental as  both the observed-data likelihood
and the usual latent-variable construction rely on each recorded observation
being an ordinary draw from the underlying mixture. In many applications,
however, the recorded observation is not collected from  a randomly selected unit. Instead, it
is measurement obtained from a unit that is believed to have the the largest value of the response variable  within a small set of candidate units, either because such
extremes arise naturally from the measurement process or because the study is
deliberately designed to retain the unit with the largest value of the response variable as  the most informative unit in each set. 

This perspective is especially appealing when ranking units within a small
randomly selected set is much cheaper or faster than taking the final
measurement on all units. Nomination-sampling is designed for such situations where inexpensive ranking is used to direct measurement effort toward
informative parts of the population \citep{willemain1980,
jozani2012randomized}.  The benefit of nomination sampling in classification problems  becomes more apparent  when the class of interest is rare. If
larger responses are more likely to arise from a rare but scientifically or
economically important subpopulation, then observing the maximum from each set
can enrich the sample for that class. In medical screening, for example,
unusually large biomarker values, lesion measurements, or repeated test
responses may be more likely to be associated with a rare diseased subgroup.
In environmental monitoring, the maximum contaminant level within a site or
time window may be the value most indicative of a rare polluted state. In
reliability or industrial studies, the largest load, stress, or degradation
measurement within a batch may be most informative about a rare high-risk
regime. Likewise, in finance or economics, one may focus on the most extreme
loss, return, or price movement among a set of assets or time periods because
such values are most informative about rare market states or tail-risk
episodes \citep{KingZeng2001,HeGarcia2009}. In all of these cases,
the observed sample is biased toward the upper tail by design or by the nature
of the measurement process, precisely because large values help direct the
study toward the subpopulation of interest.

Once observations are selected in this way, the likelihood function for the
observed sample changes, and so does the latent structure required for
EM-based inference. Consequently, treating NS observations as if they had
arisen under simple random sampling can distort both estimation and
classification. 

The purpose of this paper is to develop a fractionally supervised
classification procedure that is faithful to NS. We formulate the 
observed-data likelihood function under NS, show why the EM algorithm with standard  latent
construction breaks down, and introduce a proper latent representation that
accounts both for the class membership of the observed maximum and for the
latent composition of the remaining units in the set. This yields a valid EM
algorithm and, in turn, a coherent weighted-likelihood FSC procedure for
partially labeled nominated samples. We present the methodology in general
form, illustrate it for normal mixtures, study its finite-sample performance
through simulation, and analyze a real data set to demonstrate its practical
value. 

To our knowledge, existing FSC formulations do not incorporate nomination
sampling, and existing mixture-model work under ranked-set or nomination-type
designs \citep{hatefi2014,hatefi2020} does not address FSC. The present paper bridges this gap.

The remainder of the paper proceeds as follows. Section~\ref{sec:background}
sets up the finite mixture modeling  and FSC framework together with the NS design, establishes
the improperness of the standard single-indicator augmentation, and derives
the proper latent variables  along with the resulting E-step quantities.
Section~\ref{sec:fsc-ns} embeds this development into the FSC weighted
likelihood and derives the EM updates, illustrating both the general form and
the rare-event contamination normal-mixture model case explicitly. Section~\ref{sec:sim} reports a comprehensive simulation
study.  Section~\ref{sec:realdata} presents a real data application  and Section~\ref{sec:conclusion} closes with a
discussion of the findings and directions for future work.

%=========================================================================
\section{Preliminary results on finite mixture modeling with nominated samples}
\label{sec:background}
%=========================================================================

We first review terminologies and  notations,  and briefly describe the FSC  framework of \citet{vrbik}, and maxima nomination sampling. Throughout the paper we use the following terminology consistently.
{NS} refers to the maxima nomination sampling design in which,
for each set of $k$ candidate units, only the unit that is judged to have  the largest
value of the  variable of interest  is fully measured and retained.
{FSC-NS} denotes fractionally supervised
classification built on the correct NS weighted likelihood and proposed set of 
latent variables.
{FSC-SRS} applies the ordinary
simple-random-sampling FSC of \citet{vrbik} directly to the nominated
observations, treating them as ordinary mixture draws.  All other notation is introduced as needed.

Consider a two-component finite mixture model with density
\begin{equation}\label{eq:fmm}
f(x;\bpsi)
= \pi f_1(x;\btheta_1) + (1-\pi) f_2(x;\btheta_2),
\quad
\bpsi = (\pi,\btheta_1,\btheta_2)^\top,
\end{equation}
where $\pi\in(0,1)$ is the mixing proportion and $f_1$ and $f_2$ are
component densities indexed by $\btheta_1$ and $\btheta_2$. Let
\[
F(x;\bpsi)=\pi F_1(x;\btheta_1)+(1-\pi)F_2(x;\btheta_2)
\]
denote the corresponding mixture cdf. Although we focus on two components for
clarity, the structural arguments extend naturally to more general finite
mixtures.

In FSC applications the analyst observes three groups of data: a labeled
sample  $\{x_{1i}\}_{i=1}^{n_1}$ from component 1, a labeled sample   $\{x_{2j}\}_{j=1}^{n_2}$ from component 2, and an unlabeled
sample   $\{x_{3r}\}_{r=1}^{n_3}$ from the mixture. FSC combines these through the weighted likelihood
\begin{equation}
L_{\bw}(\bpsi\mid\bx) = \prod_{i=1}^{n_1} f^{w_1}_1(x_{1i};\btheta_1)\, \prod_{j=1}^{n_2} f^{w_2}_1(x_{2i};\btheta_2) \prod_{r=1}^{n_3}  f^{w_3}(x_{3r};\bpsi),
\end{equation}
where  $\bx= \{x_{1i}\}_{i=1}^{n_1}\cup\{x_{2j}\}_{j=1}^{n_2}\cup\{x_{3r}\}_{r=1}^{n_3}$ and 
 the weights $\bw=(w_1,w_2,w_3)^\top$ regulate the relative influence of
the three groups. Setting   $w_3=0$ yields supervised classification, whereas
$w_1=w_2=0$ gives unsupervised estimation; intermediate choices interpolate
between these extremes. To perform the usual estimation and classification with such data, one works with the log-likelihood function given by 
\begin{equation}\label{eq:wloglik-srs}
\ell_{\bw}(\bpsi\mid\bx)
= w_1\sum_{i=1}^{n_1}\log f_1(x_{1i};\btheta_1)
+ w_2\sum_{j=1}^{n_2}\log f_2(x_{2j};\btheta_2)
+ w_3\sum_{r=1}^{n_3}\log f(x_{3r};\bpsi).
\end{equation}
Note that  \eqref{eq:wloglik-srs} is the weighted version of the usual log-likelihood function of $\bx$. 

The EM algorithm provides the natural route to maximizing
\eqref{eq:wloglik-srs}. In  \citet{vrbik}, each unlabeled observation $x_{3r}$ is augmented by a
latent component indicator $Z_r\in\{0,1\}$, and the E-step computes
\begin{equation}\label{eq:estep-srs}
\tilde z_r^{(t)}
:= \E[Z_r\mid x_{3r};\bpsi^{(t)}]
= \frac{\pi^{(t)}f_1(x_{3r};\btheta_1^{(t)})}
       {f(x_{3r};\bpsi^{(t)})}.
\end{equation}
This construction is simple because, under simple random sampling, each
recorded observation corresponds to exactly one latent draw from the mixture,
so a single binary indicator captures all relevant latent information. The
M-step then separates cleanly by component.

Under NS, that structure no longer holds. Instead of observing an ordinary
mixture draw, one observes the maximum from a set of size $k$. Consequently,
the observed-data model depends not only on the mixture density but also on
the distribution of the unmeasured units in the set.

Suppose that the observed data form an NS sample of size $n$. For each
$i=1,\ldots,n$, let
$
\mathcal{U}_{i1},\ldots,\mathcal{U}_{ik}
$
be  simple random samples (sets) of  size $k$  units from the underlying distribution. The $k$ units in each set
are first ranked, typically by judgment or by means of an auxiliary variable,
without fully measuring the study variable on all set members resulting in $\mathcal{U}_{(1)i}, \ldots, \mathcal{U}_{(k)i}$. From this ordered set only  $\mathcal{U}_{(k)i}$, the unit
judged to have  the largest value of the response variable is  then fully measured and  its response value  $X_{(k)i}=\max(X_{i1},\ldots,X_{ik})$ is retained, so that  the observed sample
\[
\bX_{\NS}=\{ X_{(k)1},\ldots,X_{(k)n}\}
\]
consists of $n$ independent maxima, each obtained from a set of size $k$. If
the original draws have density $f$ and cdf $F$, then each recorded NS
observation has density
\begin{equation}\label{eq:ns-density}
g_k(x)=k\,f(x)\,F(x)^{k-1}.
\end{equation}
Thus the observed-data model involves both the density $f$ and the cdf $F$.
The recorded maximum therefore carries indirect information about the
unmeasured members of the set, and this is exactly what distinguishes the NS
likelihood from its simple random sampling counterpart.

Under the mixture model \eqref{eq:fmm}, substituting the mixture density and
cdf into \eqref{eq:ns-density} gives
\begin{equation}\label{eq:NS-density}
g_k(x;\bpsi)
=
k\,
\bigl[\pi f_1(x;\btheta_1)+(1-\pi)f_2(x;\btheta_2)\bigr]
\bigl[\pi F_1(x;\btheta_1)+(1-\pi)F_2(x;\btheta_2)\bigr]^{k-1},
\end{equation}
so the observed-data likelihood for $n$ independent NS observations is
\begin{align}
L_{\NS}(\bpsi)= L_{\NS}(\bpsi \mid \bx_{\NS})= \prod_{i=1}^n g_k(x_{(k)i};\bpsi).
%&=
%k^n\prod_{i=1}^n
%\bigl[\pi f_1(x_{(k)i};\btheta_1)+(1-\pi)f_2(x_{(k)i};\btheta_2)\bigr]
%\nonumber\\
%&\qquad\times
%\bigl[\pi F_1(x_{(k)i};\btheta_1)+(1-\pi)F_2(x_{(k)i};\btheta_2)\bigr]^{k-1}.
\label{eq:NS-lik}
\end{align}
When $k=1$, the cdf factor reduces to unity and the ordinary mixture
likelihood is recovered. For $k\ge 2$, however, the sampling design changes
the form of the likelihood in an essential way, so the standard FSC
construction cannot be transferred mechanically to the NS setting.

The issue is not merely computational. Under simple random sampling, each
unlabeled observation corresponds to a single latent mixture draw, so a
one-indicator augmentation is natural. Under NS, each recorded observation is
generated by a full set of $k$ latent draws from the mixture. The observed
maximum identifies one of those draws only indirectly, while the remaining
$k-1$ latent draws continue to affect the likelihood through the cdf term in
\eqref{eq:NS-lik}. Any EM algorithm for NS data must therefore be based on
latent variables rich enough to recover the observed-data likelihood exactly
after marginalization.

\begin{definition}[Proper latent variable]\label{def:proper}
Let $\bx$ have observed-data likelihood $L(\btheta\mid\bx)$. A latent
variable $\bZ$ with complete-data likelihood $L(\btheta,\bZ\mid\bx)$ is said
to be {proper} if
\[
\sum_{\bZ=\bz}L(\btheta,\bZ\mid\bx)=L(\btheta\mid\bx)
\]
for all $\btheta$.
\end{definition}

Under simple random sampling, the standard component indicator is proper,
since summing the complete-data likelihood over $z\in\{0,1\}$ recovers the
mixture density $\pi f_1+(1-\pi)f_2$. For NS data, however, the same
augmentation fails. A natural first attempt is to let $Z_i$ denote the
component membership of the observed maximum $x_{(k)i}$ and then proceed as
in the ordinary mixture case. This leads to
\begin{align}
L_{\mathrm{Comp, 1}}(\bpsi,\bZ\mid\bx_{\NS})
&=
k^n\prod_{i=1}^n
\pi^{kZ_i}(1-\pi)^{k(1-Z_i)}
f_1^{Z_i}f_2^{1-Z_i}
F_1^{Z_i(k-1)}F_2^{(1-Z_i)(k-1)},
\label{eq:Lcomp1}
\end{align}
where all functions are evaluated at $x_{(k)i}$. The problem is that this
construction rules out mixed sets, namely those in which the observed maximum
comes from one component while some of the remaining $k-1$ units come from
the other.

\begin{lemma}[Improperness of the single-indicator augmentation]
\label{lem:wrong}
For NS data with $k\ge 2$, the latent variable $Z_i$ leading to
\eqref{eq:Lcomp1} is not proper. Specifically,
\begin{equation}\label{eq:L1-marginal}
\sum_{\bZ}L_{\mathrm{Comp, 1}}(\bpsi,\bZ\mid\bx_{\NS})
=
k^n\prod_{i=1}^n
\Bigl[
\pi^k f_1F_1^{k-1}
+(1-\pi)^k f_2F_2^{k-1}
\Bigr]
\neq
L_{\NS}(\bpsi),
\end{equation}
where all densities and cdfs are evaluated at $x_{(k)i}$.
\end{lemma}

\begin{proof}
Summing over $Z_i\in\{0,1\}$ in the $i$th factor of
$L_{\mathrm{Comp, 1}}$ yields
\[
\pi^k f_1F_1^{k-1}+(1-\pi)^k f_2F_2^{k-1}.
\]
By contrast, the true likelihood factor in \eqref{eq:NS-lik} is
\[
(\pi f_1+(1-\pi)f_2)(\pi F_1+(1-\pi)F_2)^{k-1},
\]
whose binomial expansion contains both pure and cross-component terms. The
marginal of $L_{\mathrm{Comp, 1}}$ retains only the pure terms and therefore
fails to recover the full observed-data likelihood.
\end{proof}

\begin{figure}[h!]
\centering
\caption{Log-likelihoods $\ell_{\mathrm{C}}(\pi)$ (correct, solid blue)
and $\ell_{\mathrm{W}}(\pi)$ (misspecified, dashed red) as functions of
$\pi$ with $(\mu_2,\sigma_2)$ fixed at truth.
Components are $f_1=N(0,1)$ and $f_2=N(3.5,\,1.2^2)$, with $k=3$ and
$n=20$ NS observations generated with \texttt{set.seed(2025)}.
Both curves are shifted so their maximum is at zero.
The correct MLE $\hat\pi_C=0.54$ is close to the truth
$\pi_0=0.40$ (green dotted line); the misspecified MLE
$\hat\pi_W=0.16$ is far from it.}
\includegraphics[width=0.6\textwidth]{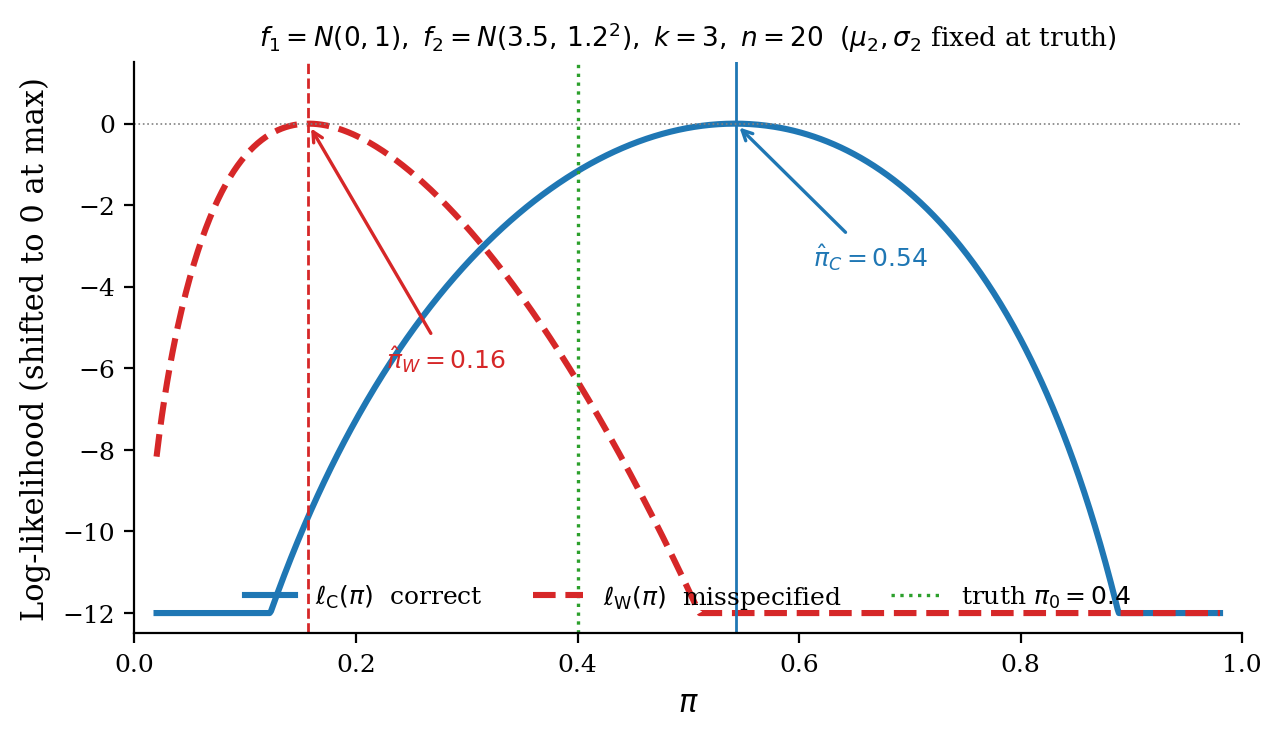}
\label{fig:lemma-2d}
\end{figure}

To make the message of Lemma~\ref{lem:wrong} concrete, we
construct a small example in which the correct log-likelihood function of NS data 
\begin{equation}\label{eq:two-objectives1}
\ell_{\mathrm{C}}(\bpsi)
=\sum_{i=1}^n\log g_k(x_{(k)i};\bpsi)
\end{equation}
and its misspecified version obtained in Lemma \ref{lem:wrong}
\begin{equation}\label{eq:two-objectives}
\ell_{\mathrm{W}}(\bpsi)
=\sum_{i=1}^n\log\!\Bigl[
\pi^k f_1(x_{(k)i};\btheta_1) F_1^{k-1}(x_{(k)i};\btheta_1)+(1-\pi)^k f_2(x_{(k)i};\btheta_2) F_2^{k-1}(x_{(k)i};\btheta_2),
\Bigr]
\end{equation}
have demonstrably different maximizers. We take $f_1=\phi(\cdot\,;0,1)$,
$f_2=\phi(\cdot\,;\mu_2,\sigma_2^2)$ with true parameters
$(\pi_0,\mu_{2,0},\sigma_{2,0})=(0.40,\,3.5,\,1.2)$, set size $k=3$,
and generate $n=20$ independent NS observations from the true
model in \textsc{R} with \texttt{set.seed(2025)}.
Figure~\ref{fig:lemma-2d} plots $\ell_{\mathrm{C}}$ and
$\ell_{\mathrm{W}}$ as functions of $\pi$ with $(\mu_2,\sigma_2)$
fixed at their true values. The two curves differ fundamentally in
shape. The correct objective $\ell_{\mathrm{C}}$ (solid blue) is
unimodal with a well-defined interior maximum at $\hat\pi_C=0.54$,
close to the truth $\pi_0=0.40$.
The misspecified objective $\ell_{\mathrm{W}}$ (dashed red) attains
its maximum at $\hat\pi_W=0.16$, far from the truth and nearly three
times smaller than $\hat\pi_C$. Maximizing the wrong objective
therefore produces a mixing proportion estimate that is both
substantially biased and in the wrong direction.

%%Figures~\ref{fig:lemma-3d} and~\ref{fig:lemma-heat} examine the

\begin{figure}[h!]
\centering
\caption{Normalized log-likelihood surfaces $\ell_{\mathrm{C}}(\mu_2,\sigma_2)$
(correct, blue) and $\ell_{\mathrm{W}}(\mu_2,\sigma_2)$ (misspecified, red)
with $\pi$ fixed at truth $\pi_0=0.40$, $f_1=N(0,1)$, $n=20$, $k=3$,
\texttt{set.seed(2025)}.
Green triangles mark the true parameter $\psi_0=(3.5,\,1.2)$.
Arrows connect $\psi_0$ to the MLE (circle) for each objective.
The correct MLE $\hat\psi_C=(3.22,\,1.19)$ lies close to the truth;
the misspecified MLE $\hat\psi_W=(2.80,\,1.27)$ is displaced by more
than $0.7$ units in $\mu_2$ and the arrow points in a qualitatively
different direction, illustrating the structural bias introduced by the
improper augmentation.}
\includegraphics[width=\textwidth]{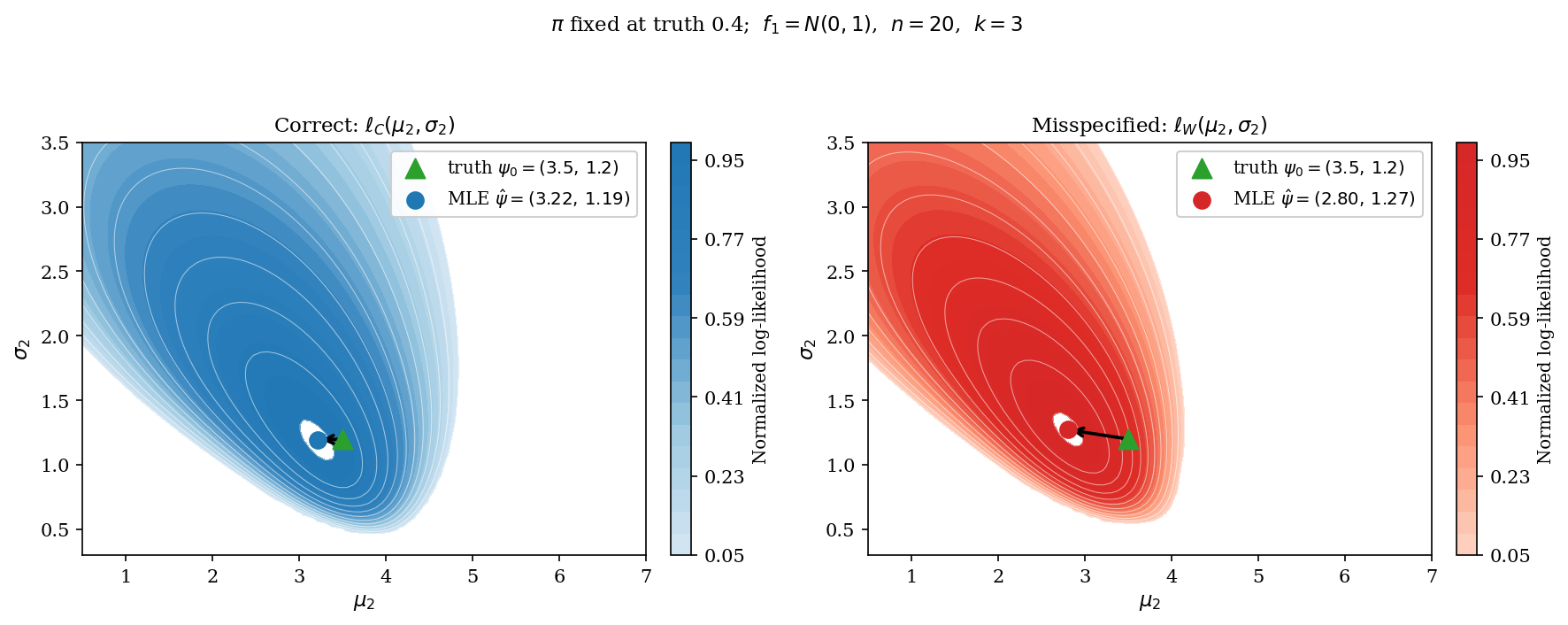}
\label{fig:lemma-heat}
\end{figure}

Figure~\ref{fig:lemma-heat} examines the
two surfaces over $(\mu_2,\sigma_2)$ with $\pi$ fixed at the truth.
Under the correct objective $\ell_{\mathrm{C}}$
(blue surface), the MLE is
$(\hat\mu_{2,C},\hat\sigma_{2,C})=(3.22,\,1.19)$,
close to the truth $(3.5,\,1.2)$.
Under the misspecified objective $\ell_{\mathrm{W}}$ (red surface),
the MLE shifts to
$(\hat\mu_{2,W},\hat\sigma_{2,W})=(2.80,\,1.27)$,
more than $0.7$ units away in $\mu_2$.

The heat maps in Figure~\ref{fig:lemma-heat} display filled contour
plots of the two log-likelihood surfaces. The arrow in each panel
connects the true parameter $\psi_0$ (green triangle) to the
corresponding MLE (coloured circle). Under $\ell_{\mathrm{C}}$ the
arrow is short and points toward the truth, indicating a small
estimation error consistent with finite-sample variability. Under
$\ell_{\mathrm{W}}$ the arrow is noticeably longer and points in a
qualitatively different direction. The misspecified surface assigns
too little weight to the right tail of the mixture, because the
cross-component cdf terms absent from $\ell_{\mathrm{W}}$ are
precisely what link the observed extremes to the second component.
Crucially, this bias is structural and cannot be removed by
increasing the sample size.

The discrepancy is directly analogous to the difference between
$(x_1+x_2)^k$ and $x_1^k+x_2^k$ as the former generates all cross terms,
whereas the latter retains only pure-power terms. The true NS likelihood has
a power-of-sums form because the $k$ latent draws within each set are i.i.d.\
from the full mixture, whereas the single-indicator construction forces a
sum-of-powers form by assigning the entire set to the component of the
observed maximum.

Because $L_{\mathrm{Comp, 1}}$ is not proper, the standard EM algorithm does
not maximize $L_{\NS}$ but a misspecified objective. In  numerical
studies, this misspecification can produce serious bias and numerical
instability when the unlabeled contribution is heavily weighted. The source of
the failure also points directly to the remedy. A single binary indicator can
encode only one piece of latent information, namely the component that
generated the observed maximum, whereas an NS observation carries two representing  the
component from which the measured extreme arose, and the component
composition of the remaining unmeasured set members.

For each NS observation $x_{(k)i}$, we therefore introduce the latent pair
$(Z_i,V_i)$. The variable $Z_i\in\{0,1\}$ records the component membership of
the observed maximum itself. The variable $V_i\in\{0,1,\ldots,k-1\}$ counts
how many of the remaining $k-1$ unmeasured set members came from
component~1. Together, $(Z_i,V_i)$ describe the component composition of the
entire set that produced $x_{(k)i}$. This type of augmented latent
representation, in which one records both the component of the observed order
statistic and the latent composition of the remaining set members, goes back
to the ranked-set-sampling mixture work of \citet{hatefi2014}. The present
paper adapts that idea to the FSC setting under NS.

The motivation for this construction comes directly from the two factors in
\eqref{eq:NS-lik}, which admit the decompositions
\begin{align}
\pi f_1(x)+(1-\pi)f_2(x)
&=
\sum_{z=0}^1
\pi^z(1-\pi)^{1-z}f_1^z(x)f_2^{1-z}(x),
\label{eq:pdf-decomp}\\[4pt]
\bigl[\pi F_1(x)+(1-\pi)F_2(x)\bigr]^{k-1}
&=
\sum_{v=0}^{k-1}\binom{k-1}{v}
\pi^v(1-\pi)^{k-1-v}
F_1^v(x)F_2^{k-1-v}(x),
\label{eq:cdf-decomp}
\end{align}
where the first corresponds to the density contribution of the observed
maximum and the second to the cdf contribution of the remaining set members.
The complete-data likelihood is therefore
\begin{align}
L_{\mathrm{Comp, 2}}(\bpsi,\bZ,\bV\mid\bx_{\NS})
=
\prod_{i=1}^n
\binom{k-1}{V_i}
\pi^{Z_i+V_i}(1-\pi)^{k-Z_i-V_i}
f_1^{Z_i}f_2^{1-Z_i}
F_1^{V_i}F_2^{k-1-V_i},
\label{eq:Lcomp2}
\end{align}
with all functions evaluated at $x_{(k)i}$.

\begin{theorem}[Proper latent variables for NS]
\label{thm:proper}
The latent pair $(Z_i,V_i)$ associated with \eqref{eq:Lcomp2} is proper,
that is,
\[
\sum_{\bZ}\sum_{\bV}
L_{\mathrm{Comp, 2}}(\bpsi,\bZ,\bV\mid\bx_{\NS})
=
L_{\NS}(\bpsi).
\]
\end{theorem}

\begin{proof}
Apply the decompositions \eqref{eq:pdf-decomp}--\eqref{eq:cdf-decomp} to
each factor of \eqref{eq:NS-lik}. Summing over the joint support of
$(Z_i,V_i)$ recovers that factor exactly, and taking the product over
$i=1,\ldots,n$ yields $L_{\NS}(\bpsi)$.
\end{proof}

%\begin{proposition}[Identifiability under NS]
%\label{prop:ident}
%Suppose the underlying mixture $f(\cdot;\bpsi)=\pi f_1(\cdot;\btheta_1)
%+(1-\pi)f_2(\cdot;\btheta_2)$ is identifiable in the sense that
%$f(\cdot;\bpsi)=f(\cdot;\bpsi')$ almost everywhere implies $\bpsi=\bpsi'$.
%Then the NS density $g_k(\cdot;\bpsi)=k\,f(\cdot;\bpsi)\,F(\cdot;\bpsi)^{k-1}$
%is also identifiable in the same sense.
%\end{proposition}
%
%\begin{proof}
%Because $g_k(\cdot;\bpsi)$ is a proper density on $\mathbb{R}$, it
%uniquely determines its own cdf
%$G_k(x;\bpsi)=\int_{-\infty}^x g_k(t;\bpsi)\,dt = F(x;\bpsi)^k$.
%If $g_k(\cdot;\bpsi)=g_k(\cdot;\bpsi')$ almost everywhere, then
%$G_k(\cdot;\bpsi)=G_k(\cdot;\bpsi')$ everywhere, so
%$F(\cdot;\bpsi)^k=F(\cdot;\bpsi')^k$.
%Since $F(\cdot;\bpsi)$ and $F(\cdot;\bpsi')$ are cdfs taking values in
%$[0,1]$, the map $u\mapsto u^k$ is strictly monotone on $[0,1]$, and
%therefore $F(\cdot;\bpsi)=F(\cdot;\bpsi')$ everywhere.
%Differentiating gives $f(\cdot;\bpsi)=f(\cdot;\bpsi')$ almost everywhere,
%and identifiability of the underlying mixture then yields $\bpsi=\bpsi'$.
%\end{proof}
%
%\begin{remark}
%The key observation is that $G_k(x;\bpsi)=F(x;\bpsi)^k$, which is the cdf
%of the maximum of $k$ i.i.d.\ draws from $F(\cdot;\bpsi)$.  This relation
%allows one to move directly from equality of the NS density to equality of
%the mixture cdf, bypassing any need to work with the product $fF^{k-1}$
%directly.
%\end{remark}

With the latent pair $(Z_i,V_i)$ in place, the E-step requires the
conditional expectations of $Z_i$ and $V_i$ given the observed nominated
value under the current parameter estimates.

\begin{proposition}[E-step quantities]
\label{prop:estep}
Under the proper complete-data model \eqref{eq:Lcomp2}, the E-step
sufficient statistics for NS data are
\begin{align}
\tilde z_i^{(t)}
&:=
\E[Z_i\mid x_{(k)i};\bpsi^{(t)}]
=
\frac{\pi^{(t)}f_1(x_{(k)i};\btheta_1^{(t)})}
     {f(x_{(k)i};\bpsi^{(t)})},
\label{eq:Ez-NS}\\[6pt]
\tilde v_i^{(t)}
&:=
\E[V_i\mid x_{(k)i};\bpsi^{(t)}]
=
(k-1)\,
\frac{\pi^{(t)}F_1(x_{(k)i};\btheta_1^{(t)})}
     {F(x_{(k)i};\bpsi^{(t)})}.
\label{eq:Ev-NS}
\end{align}
\end{proposition}

\begin{proof}
Summing the complete-data posterior over $V_i$ yields
\[
P(Z_i=1\mid x_{(k)i})
=
\frac{\pi f_1(x_{(k)i}; \btheta_1)}{f(x_{(k)i};\bpsi)},
\]
which gives \eqref{eq:Ez-NS}. Conditional on $x_{(k)i}$, the variable
$V_i$ has a binomial distribution with parameters $k-1$ and success
probability $\pi F_1(x_{(k)i};\btheta_1)/F(x_{(k)i};\bpsi)$, whose mean gives
\eqref{eq:Ev-NS}.
\end{proof}

The form of \eqref{eq:Ez-NS} and \eqref{eq:Ev-NS} highlights the
essential difference between ordinary random sampling and NS. The quantity
$\tilde z_i^{(t)}$ has the same algebraic form as the usual posterior
component probability, whereas $\tilde v_i^{(t)}$ is governed by the cdf
rather than the density. Together, these two quantities capture the two
sources of information carried by an NS observation contained in  the observed maximum
itself and the latent contribution of the unmeasured set members.

\section{Fractionally supervised classification with nominated samples}
\label{sec:fsc-ns}
%=========================================================================

We now embed the proper EM formulation for nomination sampling within FSC framework. 
%The key observation is
%that under NS with set size $k$, the labeled samples are themselves nominated
%observations from their respective component distributions, while the unlabeled
%sample consists of nominated observations from the mixture.
%
Let
\[
\bY_1=(Y_{11},\ldots,Y_{1n_1})^\top,\qquad
\bY_2=(Y_{21},\ldots,Y_{2n_2})^\top,\qquad
\bY_3=(Y_{31},\ldots,Y_{3n_3})^\top
\]
denote, respectively, the labeled NS sample from component 1, the labeled NS sample
from component 2, and the unlabeled NS sample from the mixture. Under NS,
each $Y_{1i}$ is the maximum of $k$ draws from $f_1$, each $Y_{2j}$ is the
maximum of $k$ draws from $f_2$, and each $Y_{3r}$ is the maximum of $k$ draws
from the mixture density $f$. The corresponding observed-data densities are
\[
g_1(y;\btheta_1)=k f_1(y;\btheta_1)F_1(y;\btheta_1)^{k-1},\qquad
g_2(y;\btheta_2)=k f_2(y;\btheta_2)F_2(y;\btheta_2)^{k-1},
\]
and
\[
g_3(y;\bpsi)=k f(y;\bpsi)F(y;\bpsi)^{k-1}.
\]
Given nonnegative weights $\bw=(w_1,w_2,w_3)^\top$, the FSC-NS weighted
likelihood function is 
\begin{equation}
L_{\bw}(\bpsi\mid\bX) = \prod_{i=1}^{n_1} g^{w_1}_1(y_{1i};\btheta_1)\, \prod_{j=1}^{n_2} g^{w_2}_1(y_{2i};\btheta_2) \prod_{r=1}^{n_3}  g^{w_3}(y_{3r};\bpsi),
\end{equation}
which results in the following FSC-NS weighted
log-likelihood function %
\begin{align}
\ell_{\bw}(\bpsi\mid\bY)
&=
w_1\sum_{i=1}^{n_1}
\Bigl[\log f_1(y_{1i};\btheta_1)+(k-1)\log F_1(y_{1i};\btheta_1)\Bigr]
\nonumber\\
&\quad+
w_2\sum_{j=1}^{n_2}
\Bigl[\log f_2(y_{2j};\btheta_2)+(k-1)\log F_2(y_{2j};\btheta_2)\Bigr]
\nonumber\\
&\quad+
w_3\sum_{r=1}^{n_3}
\log\Bigl[k\,f(y_{3r};\bpsi)\,F(y_{3r};\bpsi)^{k-1}\Bigr].
\label{eq:wloglik-ns}
\end{align}
When $k=1$, the cdf contributions vanish and \eqref{eq:wloglik-ns} reduces to
the usual FSC-SRS weighted likelihood in \eqref{eq:wloglik-srs}. To maximize \eqref{eq:wloglik-ns}, we introduce for each unlabeled
observation $Y_{3r}$ the proper latent pair $(Z_r,V_r)$ and take the
conditional expectation of the weighted complete-data log-likelihood. This
gives the Q-function
\begin{align}
Q(\bpsi;\bpsi^{(t)})
&=
w_1\sum_{i=1}^{n_1}
\Bigl[\log f_1(y_{1i};\btheta_1)+(k-1)\log F_1(y_{1i};\btheta_1)\Bigr]
\nonumber\\
&\quad+
w_2\sum_{j=1}^{n_2}
\Bigl[\log f_2(y_{2j};\btheta_2)+(k-1)\log F_2(y_{2j};\btheta_2)\Bigr]
\nonumber\\
&\quad+
w_3\sum_{r=1}^{n_3}
\Bigl[
(\tilde z_r+\tilde v_r)\log\pi
+(k-\tilde z_r-\tilde v_r)\log(1-\pi)
\nonumber\\
&\qquad\qquad\qquad
+\tilde z_r\log f_1(y_{3r};\btheta_1)
+(1-\tilde z_r)\log f_2(y_{3r};\btheta_2)
\nonumber\\
&\qquad\qquad\qquad
+\tilde v_r\log F_1(y_{3r};\btheta_1)
+(k-1-\tilde v_r)\log F_2(y_{3r};\btheta_2)
\Bigr],
\label{eq:Q-ns}
\end{align}
where $\tilde z_r=\tilde z_r^{(t)}$ and $\tilde v_r=\tilde v_r^{(t)}$ are
the E-step quantities from Proposition~\ref{prop:estep}.

The structure of \eqref{eq:Q-ns} is especially revealing. The first two lines
are the weighted NS log-likelihood contributions of the labeled groups. The
unlabeled part then separates into three components consisting of a mixing-proportion term,
a density-based soft-classification term, and a cdf-based term. The last of
these has no SRS analogue and is precisely where the nomination mechanism
enters estimation. It transfers information from the unmeasured members of
each set into the component updates.

Maximizing \eqref{eq:Q-ns} with respect to $\pi$ gives
\begin{equation}\label{eq:pi-update-ns}
\pi^{(t+1)}
=
\frac{\sum_{r=1}^{n_3}(\tilde z_r+\tilde v_r)}{n_3k}.
\end{equation}
The denominator is $n_3k$, not $n_3$, because $Z_r+V_r$ counts the total
number of latent component-1 draws among the $k$ observations that produced
$Y_{3r}$. Using $n_3$ instead would implicitly treat each nominated
observation as if it represented only one latent draw, which is exactly the
SRS logic that fails under nomination sampling.

The component updates are
\begin{align}
\btheta_1^{(t+1)}
&=
\arg\max_{\btheta_1}
\Biggl\{
w_1\sum_{i=1}^{n_1}
\Bigl[\log f_1(y_{1i};\btheta_1)+(k-1)\log F_1(y_{1i};\btheta_1)\Bigr]
\nonumber\\
&\qquad\qquad\qquad\qquad
+w_3\sum_{r=1}^{n_3}
\Bigl[\tilde z_r\log f_1(y_{3r};\btheta_1)
+\tilde v_r\log F_1(y_{3r};\btheta_1)\Bigr]
\Biggr\},
\label{eq:theta1-ns}\\[5pt]
\btheta_2^{(t+1)}
&=
\arg\max_{\btheta_2}
\Biggl\{
w_2\sum_{j=1}^{n_2}
\Bigl[\log f_2(y_{2j};\btheta_2)+(k-1)\log F_2(y_{2j};\btheta_2)\Bigr]
\nonumber\\
&\qquad\qquad\qquad\qquad
+w_3\sum_{r=1}^{n_3}
\Bigl[(1-\tilde z_r)\log f_2(y_{3r};\btheta_2)
+(k-1-\tilde v_r)\log F_2(y_{3r};\btheta_2)\Bigr]
\Biggr\}.
\label{eq:theta2-ns}
\end{align}
These expressions make the NS correction transparent. The observed maximum is
soft-allocated through $\tilde z_r$, while the unmeasured members of the set
enter through the cdf weights $\tilde v_r$.

At convergence, classification of $Y_{3r}$ is based on the posterior
probability that the observed maximum came from component 1. Equivalently, the
Bayes rule assigns $Y_{3r}$ to component 1 whenever
\[
\frac{f_1(y_{3r};\hat\btheta_1)}{f_2(y_{3r};\hat\btheta_2)}
>
\frac{1-\hat\pi}{\hat\pi}.
\]
Thus the classification rule has the same formal structure as in the ordinary
mixture case; what changes under NS is the estimation step.

In practice, we initialize the EM algorithm using a clustering rule on the
pooled sample $(\bY_1,\bY_2,\bY_3)$, such as $k$-means applied to the observed
nominated values. Cluster proportions provide initial values for $\pi$, and
within-cluster moment- or likelihood-based estimates provide starting values
for the component parameters. In our numerical studies, convergence is
declared when
\[
\bigl|\ell_{\bw}(\bpsi^{(t+1)}\mid\bY)-\ell_{\bw}(\bpsi^{(t)}\mid\bY)\bigr|
\le 10^{-5}.
\]

The preceding development is fully general. We next give two examples that
will also serve as the basis for the simulation studies.

\begin{example}[Normal mixture]
Suppose
\[
f(x;\bpsi)=\pi f_1(x;\btheta_1)+(1-\pi)f_2(x;\btheta_2),
\qquad
f_m(\cdot;\btheta_m)=\phi(\cdot;\mu_m,\sigma_m^2),\quad m=1,2,
\]
where $\phi$ and $\Phi$ denote the normal density and cdf, and
$\btheta_m=(\mu_m,\sigma_m^2)$. In this case, the density-based part of the
M-step retains the familiar Gaussian soft-assignment form. Writing
$e_{1r}=\tilde z_r$, $e_{2r}=1-\tilde z_r$, and
\[
\tilde n_m = w_m n_m + w_3\sum_{r=1}^{n_3} e_{mr},
\]
the corresponding updates are
\begin{equation}\label{eq:mu-update}
\mu_m^{(t+1)}
=
\frac{
w_m\sum_i^{n_m} y_{mi}
+
w_3\sum_r ^{n_3}e_{mr}y_{3r}
}{
\tilde n_m
},
\qquad m=1,2,
\end{equation}
and
\begin{equation}\label{eq:sigma-update}
(\sigma_m^2)^{(t+1)}
=
\frac{
w_m\sum_i^{n_m} (y_{mi}-\mu_m^{(t)})^2
+
w_3\sum_r^{n_3} e_{mr}(y_{3r}-\mu_m^{(t)})^2
}{
\tilde n_m
}.
\end{equation}
These updates account for the density contribution in closed form and provide
natural starting values for the full M-step. The NS correction enters through
the cdf terms
\[
\tilde v_r\log \Phi(y_{3r};\mu_m,\sigma_m),
\]
which have no SRS analogue. Consequently, the full update is not available in
closed form, but can be obtained efficiently by a low-dimensional numerical
refinement, for example a Newton step initialized at
\eqref{eq:mu-update}--\eqref{eq:sigma-update}. When $k=1$, the $\tilde v_r$
terms vanish and the ordinary Gaussian FSC-SRS updates are recovered.
\end{example}

The next example focuses on a setting in which one component is rare. This is
relevant in applications such as environmental monitoring, medical screening,
fraud detection, and reliability studies, where the scientifically important
class may constitute only a small fraction of the population. The model may
also be viewed as a one-sided contamination mixture, linking the FSC-NS
framework to the classical contamination literature and to modern work on
rare-event and imbalanced classification. See 
\citet{KingZeng2001} for rare-events methodology, and
\citet{HeGarcia2009} for a broader view of imbalanced classification.

\begin{example}[Rare-event contamination mixture]
Consider
\begin{equation}\label{eq:rare-mixture}
X \sim (1-\varepsilon)\,N(0,1)+\varepsilon\,N(\delta,\tau^2),
\end{equation}
where the background component $f_1$ is $N(0,1)$ and the rare signal
component $f_2$ is $N(\delta,\tau^2)$. Here $\varepsilon$ is the rare-class
proportion, assumed to be small, $\delta$ controls separation between the two
components, and $\tau$ controls the spread of the rare component. The
background parameters are treated as known, so the unknown parameter vector is
$\bpsi=(\varepsilon,\delta,\tau)$.

Under NS, the weighted log-likelihood becomes
\begin{align}
\ell_{\bw}(\bpsi\mid\bY)
&= w_1 \sum_{i=1}^{n_1}
   \Bigl[\log\phi(y_{1i})+(k-1)\log\Phi(y_{1i})\Bigr]
\nonumber\\
&\quad + w_2\sum_{j=1}^{n_2}
   \Bigl[\log\phi(y_{2j};\delta,\tau)+(k-1)\log\Phi(y_{2j};\delta,\tau)\Bigr]
\nonumber\\
&\quad + w_3\sum_{r=1}^{n_3}\log\bigl[g(y_{3r};\bpsi)\bigr],
\label{eq:wloglik-rare}
\end{align}
where $\phi(\cdot)=\phi(\cdot;0,1)$ denotes the background density and
\[
g(y;\bpsi)
=
k\Bigl[(1-\varepsilon)\phi(y)+\varepsilon\phi(y;\delta,\tau)\Bigr]
\Bigl[(1-\varepsilon)\Phi(y)+\varepsilon\Phi(y;\delta,\tau)\Bigr]^{k-1}
\]
is the NS density induced by the underlying population mixture.

The E-step again uses the latent pair $(Z_r,V_r)$, with
\begin{align}
\tilde z_r^{(t)}
&=
\frac{\varepsilon^{(t)}\phi(y_{3r};\delta^{(t)},\tau^{(t)})}
     {(1-\varepsilon^{(t)})\phi(y_{3r})
      +\varepsilon^{(t)}\phi(y_{3r};\delta^{(t)},\tau^{(t)})},
\label{eq:rare-ztilde}\\[4pt]
\tilde v_r^{(t)}
&=
(k-1)\,
\frac{\varepsilon^{(t)}\Phi(y_{3r};\delta^{(t)},\tau^{(t)})}
     {(1-\varepsilon^{(t)})\Phi(y_{3r})
      +\varepsilon^{(t)}\Phi(y_{3r};\delta^{(t)},\tau^{(t)})}.
\label{eq:rare-vtilde}
\end{align}
The mixing-proportion update is therefore
\begin{equation}\label{eq:rare-eps-update}
\varepsilon^{(t+1)}
=
\frac{\sum_{r=1}^{n_3}\bigl(\tilde z_r^{(t)}+\tilde v_r^{(t)}\bigr)}
     {n_3k},
\end{equation}
which is the direct specialization of \eqref{eq:pi-update-ns}.

Since the background component is known, only $(\delta,\tau)$ must be updated.
Their M-step is obtained by maximizing
\begin{align}
(\delta^{(t+1)},\tau^{(t+1)})
&=
\arg\max_{\delta,\tau>0}
\Biggl\{
w_2\sum_{j=1}^{n_2}
\Bigl[\log\phi(y_{2j};\delta,\tau)
+(k-1)\log\Phi(y_{2j};\delta,\tau)\Bigr]
\nonumber\\
&\qquad\qquad
+w_3\sum_{r=1}^{n_3}
\Bigl[\tilde z_r^{(t)}\log\phi(y_{3r};\delta,\tau)
+\tilde v_r^{(t)}\log\Phi(y_{3r};\delta,\tau)\Bigr]
\Biggr\}.
\label{eq:rare-delta-tau-update}
\end{align}
Because the cdf terms depend nonlinearly on both $\delta$ and $\tau$, this
step does not admit a closed form. In practice, one can optimize jointly over
$\delta$ and $\log\tau$ to enforce positivity.

At convergence, the posterior probability that the observed maximum in set
$r$ originated from the rare component is
\begin{equation}\label{eq:rare-posterior}
\hat p_r
=
\frac{\hat\varepsilon\,\phi(y_{3r};\hat\delta,\hat\tau)}
     {(1-\hat\varepsilon)\phi(y_{3r})
      +\hat\varepsilon\,\phi(y_{3r};\hat\delta,\hat\tau)}.
\end{equation}
Thus $Y_{3r}$ may be assigned to the rare class whenever $\hat p_r>0.5$, or
more generally via a threshold chosen to reflect the application-specific
trade-off between false positives and missed detections.
\end{example}

\begin{remark}
These examples illustrate the main message of this section. The FSC framework
persists under nomination sampling, but the EM algorithm must be modified to
account for the latent information carried by the unmeasured units within each
set. Once this correction is incorporated through the latent pair $(Z_r,V_r)$,
the method reduces to the ordinary FSC formulation when $k=1$ and extends
naturally to a broad class of parametric mixture models. Extensive simulation
studies and data analyses reported in \citet{Wang2020MSc} show that working with misspecified log-likelihood function can
substantially reduce the accuracy of parameter estimation and can adversely
affect classification performance. For this reason, in all simulation studies and real-data applications, we
report results only under the correct FSC--NS formulation and compare them with
analyses that ignore the rank information and treat the nomination-sampling
data as if they were obtained by simple random sampling, a baseline that is
still commonly used in practice (e.g., \citet{Wang2025}).
\end{remark}

\section{Simulation Studies}
\label{sec:sim}

We study the rare-event contamination mixture
\begin{equation}\label{eq:rare-mixture}
X \;\sim\; (1-\varepsilon)\,N(0,1) \;+\; \varepsilon\,N(\delta,\tau^2),
\end{equation}
where $N(0,1)$ is the majority component and $N(\delta,\tau^2)$ is the rare
signal component. The background parameters are known; the unknowns are
$\bpsi=(\varepsilon,\delta,\tau)$. We use the parameter grid
\[
\varepsilon\in\{0.02,0.05,0.10\},\quad
\delta\in\{3,4,5\},\quad
\tau\in\{1,1.5,2\},\quad
k\in\{2,3,5,8\},
\]
with labeled sizes $n_0=20$ (majority) and $n_1=10$ (rare), unlabeled
size $n_3\in\{100,200\}$, weight grid $w_3\in\{0,1,\ldots,10\}$, and
$B=500$ Monte Carlo replications. We compare FSC-NS, which uses the
correct NS weighted likelihood with latent pair $(Z_r,V_r)$, against
FSC-SRS, which treats nominated observations as ordinary mixture draws.
Additional studies for various  balanced-mixture settings can be found  in \citet{Wang2020MSc}.

\medskip
\noindent\textbf{Dell-Clutter imperfect ranking:}
To assess robustness to ranking error we generate the unlabeled data under
the Dell-Clutter model \citep{dell1972}. For each candidate unit with
study variable $Y$ we construct a ranking score $Q$, with 
\begin{equation}\label{eq:dc}
Q \;=\; \rho\,\widetilde Y \;+\; \sqrt{1-\rho^2}\;E,
\qquad
\widetilde Y = \frac{Y-\mu_Y}{\sigma_Y},
\quad E\sim N(0,1)\text{ independent of }Y,
\end{equation}
and nominate the unit with the largest $Q$ in each set of size $k$.
The correlation $\rho=\mathrm{Corr}(Q,\widetilde Y)$ indexes ranking
quality: $\rho=1$ is perfect ranking and smaller $\rho$ introduces more
ordering error. We study $\rho\in\{1,0.85,0.60\}$, where $\rho=0.85$
represents mild ranking error, realistic in many field settings where an
inexpensive auxiliary measurement is used to rank units. 

\medskip
\noindent\textbf{Performance criteria:}
For rare-event problems, overall misclassification rate alone is misleading;
a classifier that labels every observation as majority achieves low error
while missing the signal class entirely. We therefore report   the following  classification metrics 
\begin{align*}
\text{Sensitivity} &= \frac{\mathrm{TP}}{\mathrm{TP}+\mathrm{FN}},\qquad
\text{Specificity}  = \frac{\mathrm{TN}}{\mathrm{TN}+\mathrm{FP}},\qquad
\text{Precision}    = \frac{\mathrm{TP}}{\mathrm{TP}+\mathrm{FP}},\\[4pt]
F_1 &= \frac{2\,\text{Precision}\times\text{Sensitivity}}
            {\text{Precision}+\text{Sensitivity}},\qquad
\text{Bal.Acc.} = \frac{\text{Sensitivity}+\text{Specificity}}{2},
\end{align*}
where TP, FP, TN, FN denote the counts for confusion matrix associated with 
the rare class.The adjusted Rand index $\mathrm{ARI}\in[-1,1]$ measures overall partition
recovery, with $\mathrm{ARI}=1$ for perfect agreement and $\mathrm{ARI}
\approx0$ for chance-level assignment \citep{hubert(1985),steinley(2004), vrbik}. For ARI, $F_1$, Sensitivity,
Specificity, Precision, Balanced Accuracy, and AUC, {higher is
better}. For parameter estimation,
\[
\widehat{\mathrm{Bias}}(\hat\theta)=B^{-1}\!\sum_{b=1}^B(\hat\theta_b-\theta),
\quad
\widehat{\mathrm{RMSE}}(\hat\theta)=\bigl\{B^{-1}\!\sum_{b=1}^B(\hat\theta_b-\theta)^2\bigr\}^{1/2},
\quad\theta\in\{\varepsilon,\delta,\tau\},
\]
where {lower RMSE is better}. The EM algorithm is initialized by
$k$-means on the pooled sample and stopped when
$|\ell_{\bw}^{(t+1)}-\ell_{\bw}^{(t)}|\le10^{-5}$, with a ceiling of 500 iterations.

NS concentrates the unlabeled sample toward the right tail where the rare
component is most prominent. We quantify this through the enrichment ratio
$\mathrm{ER}(k)$, the ratio of the posterior rare-class probability for a
nominated maximum to the prior $\varepsilon$. At the reference scenario
$(\varepsilon,\delta,\tau)=(0.05,4,1.5)$, numerical evaluation gives
$\mathrm{ER}(2)=2.3$, $\mathrm{ER}(3)=3.1$, $\mathrm{ER}(5)=4.6$, and
$\mathrm{ER}(8)=6.2$, so the mechanism multiplies the effective signal
prevalence by two to six depending on $k$. This enrichment is valuable only
when the likelihood correctly incorporates the sampling design.

\medskip
\noindent\textbf{Overview of results:}
Table~\ref{tab:overview} provides a concise summary of average performance
across the full 27-cell $(\varepsilon,\delta,\tau)$ grid at $\rho=0.85$,
$n_3=200$, $w_3=3$, for each set size $k$.  It is intended to orient the
reader before the detailed results below.  FSC-NS delivers stable ARI and
$F_1$ across all $k$, while FSC-SRS degrades sharply as $k$ grows and
enrichment intensifies.

\begin{table}[tbp]
\centering
\caption{Overview of average classification performance at $\rho=0.85$,
$n_3=200$, $w_3=3$, averaged over the full 27-cell $(\varepsilon,\delta,\tau)$
grid ($B=500$ replications). ARI and $F_1$ are higher-is-better.
The sharpening FSC-SRS degradation with $k$ reflects growing enrichment
under the misspecified likelihood.}
\label{tab:overview}
\setlength{\tabcolsep}{6pt}
\begin{tabular}{@{}c rr rr@{}}
\toprule
& \multicolumn{2}{c}{{FSC-NS}} & \multicolumn{2}{c}{{FSC-SRS}} \\
\cmidrule(lr){2-3}\cmidrule(lr){4-5}
$k$ & ARI$\uparrow$ & $F_1\uparrow$ & ARI$\uparrow$ & $F_1\uparrow$ \\
\midrule
2 & 0.805 & 0.847 & 0.678 & 0.759 \\
3 & 0.803 & 0.859 & 0.391 & 0.585 \\
5 & 0.792 & 0.876 & 0.084 & 0.443 \\
8 & 0.777 & 0.891 & 0.010 & 0.472 \\
\bottomrule
\end{tabular}
\end{table}

\begin{figure}[tbp]
\centering
\begin{subfigure}{\textwidth}
    \centering
    \includegraphics[width=\textwidth]{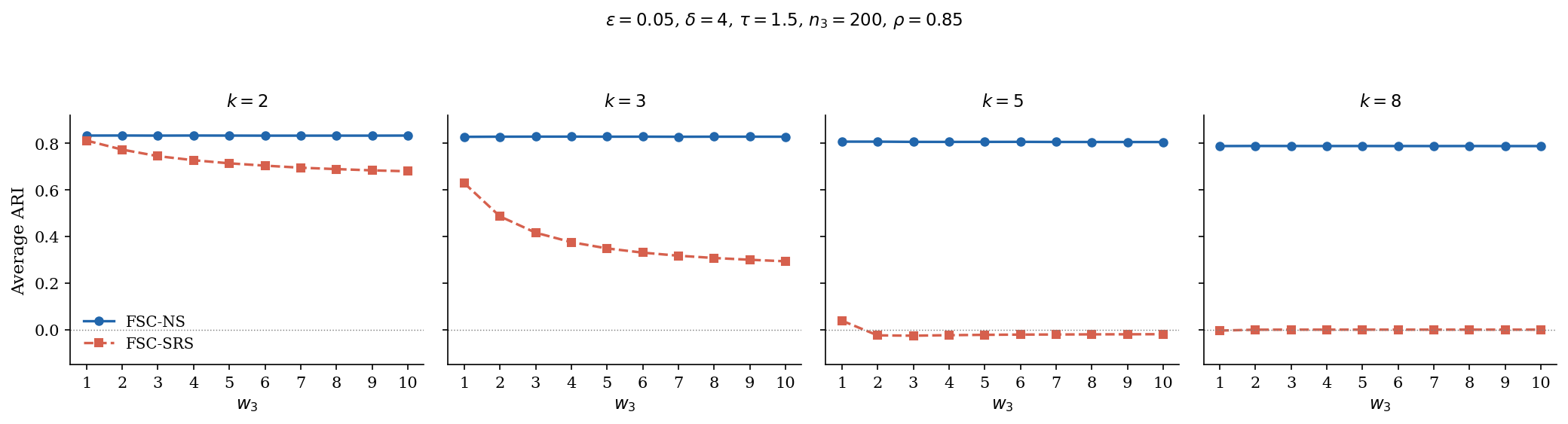}
    \caption{Average ARI versus $w_3$.}
\end{subfigure}

\vspace{0.75em}

\begin{subfigure}{\textwidth}
    \centering
    \includegraphics[width=\textwidth]{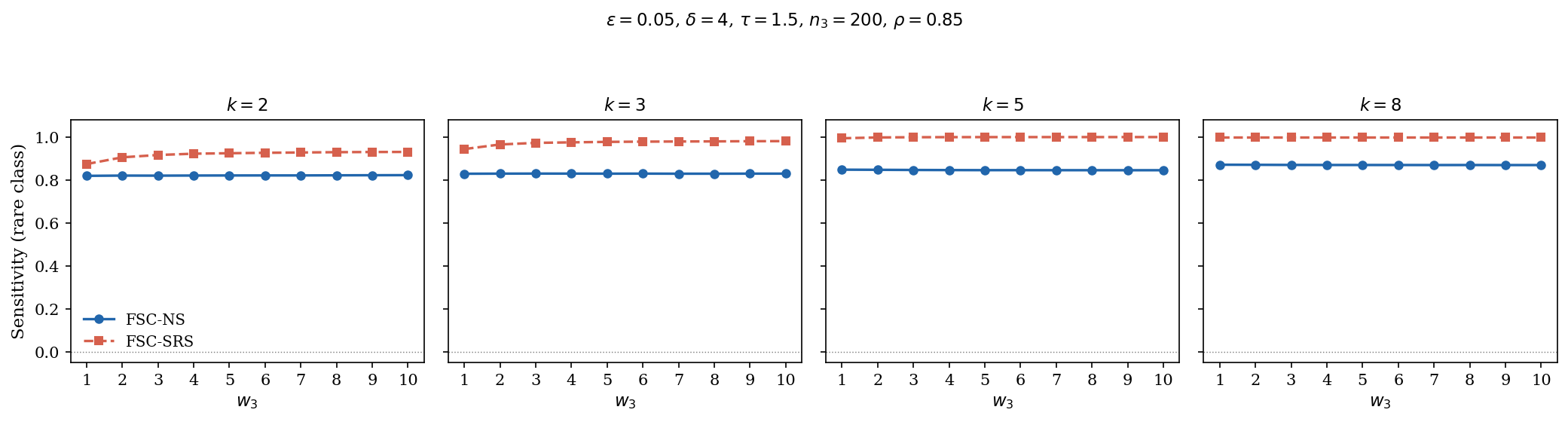}
    \caption{Sensitivity for the rare class versus $w_3$.}
\end{subfigure}

\caption{Performance versus $w_3$ at $\rho=0.85$, $\varepsilon=0.05$,
$\delta=4$, $\tau=1.5$, and $n_3=200$, for set sizes $k\in\{2,3,5,8\}$.
Panel (a) shows average ARI. FSC--NS (solid, circles) remains close to $0.83$
across all $k$ and weights, whereas FSC--SRS (dashed, squares) deteriorates sharply
for $k\ge 3$ as $w_3$ increases, with ARI approaching zero for $k=5$ and $k=8$
even at moderate weights. Panel (b) shows sensitivity for the rare class.
The near-perfect sensitivity of FSC--SRS for $k\ge 3$ is degenerate, reflecting
its tendency to label almost all observations as rare, as confirmed by the near-zero
ARI and collapsed specificity. In contrast, FSC--NS maintains calibrated sensitivity
near $0.83$ with specificity above $0.99$.}
\label{fig:rare-ari-sens}
\end{figure}

\begin{figure}[tbp]
\centering
\caption{Average rare-class $F_1$ score at $\rho=0.85$, $\tau=1.5$,
$k=3$, $n_3=200$, $w_3=3$ as a function of $(\varepsilon,\delta)$.
$F_1\in[0,1]$; higher is better. Both panels share the same colour scale.
FSC-NS achieves $F_1\ge0.67$ in every cell; FSC-SRS falls to
$0.13$ in the hardest setting $(\varepsilon=0.02,\delta=3)$. The gap is
largest at small $\varepsilon$ and moderate $\delta$, where NS enrichment
is greatest and misspecification most damaging.}
\includegraphics[width=\textwidth]{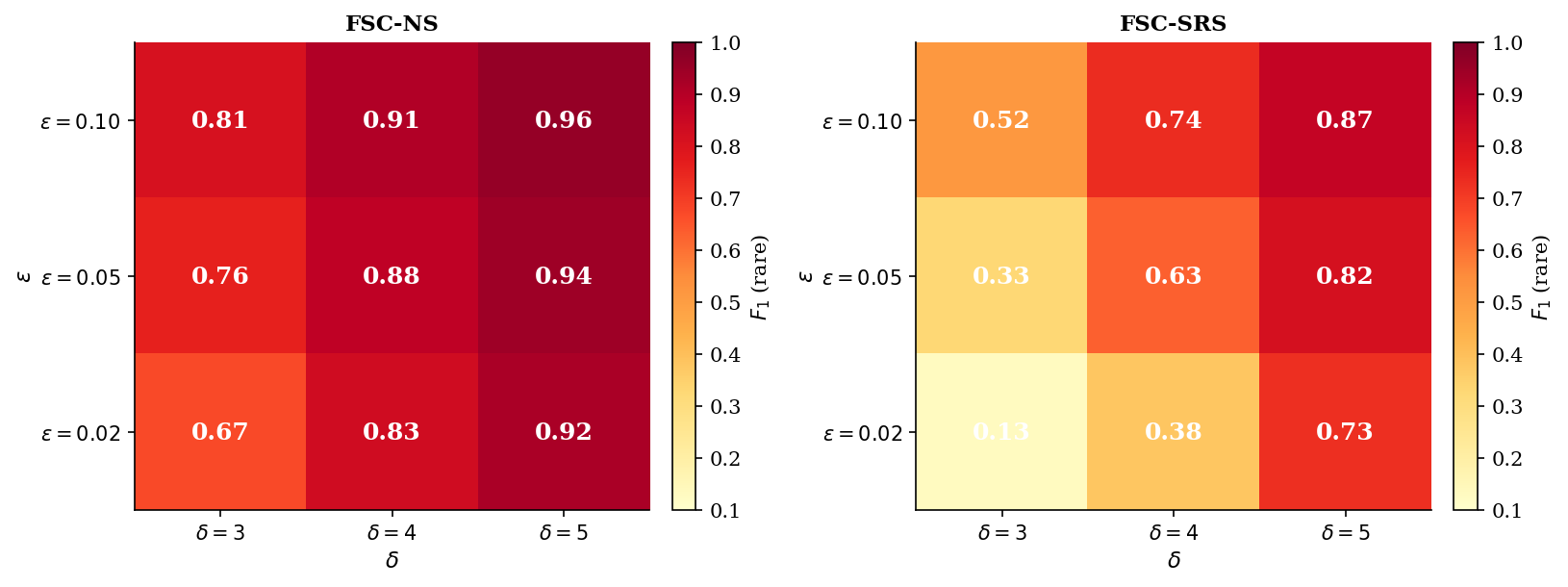}
\label{fig:rare-f1-heat}
\end{figure}

\medskip
\noindent\textbf{Classification performance:}
Tables~\ref{tab:rare-classif} and~\ref{tab:rare-k-effect} report detailed
results at the reference configuration $n_3=200$, $k=3$, $w_3=3$, and
ranking level $\rho=0.85$ (mild ranking error). Results for  $\rho=1$ and $\rho=0.60$ patterns are summarized in
Table~\ref{tab:rare-imperfect-classif}.
Under $\rho=0.85$, FSC-NS achieves $\overline{\mathrm{ARI}}$ between
$0.63$ and $0.96$ and $\overline{F_1}$ between $0.67$ and $0.98$ across the
full 27-cell grid (Table~\ref{tab:rare-classif}). Both metrics improve
monotonically with $\delta$ and with $\varepsilon$, and show that  larger separation and
higher prevalence make the rare class more identifiable. Specificity exceeds
$0.96$ in every configuration, confirming that the high sensitivity is genuine
and not achieved by over-predicting the rare class.
FSC-SRS again tells a  different story. For $\varepsilon\le0.05$ and
$\delta\le4$, its $\overline{\mathrm{ARI}}$ is near zero even though
sensitivity is near one. In other words, FSC-SRS produces  a degenerate classifier that
labels almost every observation as rare. The degenerate sensitivity is not
informative; it arises because the misspecified mixing-proportion update uses
denominator $n_3$ instead of $n_3k$, inflating $\hat\varepsilon$ toward one.
The $\tilde v_r$ correction in FSC-NS's M-step prevents this entirely.
Figure~\ref{fig:rare-ari-sens}(a) plots average ARI against $w_3$ at $\rho=0.85$
for each $k$. FSC-NS remains near $0.83$ across the full weight grid
at every $k$. FSC-SRS deteriorates sharply with $w_3$ for $k\ge3$, with ARI near
zero at $k=5$ or $k=8$ even for moderate weights. Figure~\ref{fig:rare-ari-sens}(b)  shows that FSC-SRS
attains near-perfect sensitivity at $k\ge3$ only through degenerate all-rare
prediction, whereas FSC-NS maintains calibrated sensitivity near $0.83$
with specificity above $0.99$.
Table~\ref{tab:rare-k-effect} isolates the effect of set size at $\rho=0.85$.
Under FSC-NS, ARI moves from $0.834$ at $k=2$ to $0.789$ at $k=8$
while $F_1$ improves from $0.869$ to $0.908$, because larger sets produce more
informative nominated maxima. Under FSC-SRS, ARI drops to near zero for $k\ge5$
even though sensitivity reaches one. Figure~\ref{fig:rare-f1-heat} plots the rare-class $F_1$ at $\rho=0.85$ as a
function of $(\varepsilon,\delta)$ at $\tau=1.5$, $k=3$, $n_3=200$, $w_3=3$.
Under FSC-NS, $F_1\ge0.67$ in every cell; under FSC-SRS it falls to
$0.13$ in the hardest setting $(\varepsilon=0.02,\delta=3)$. The gap is
largest at small $\varepsilon$ and moderate $\delta$, the regime where
enrichment is greatest and misspecification most damaging.
\begin{table}[tbp]
\centering
\caption{Classification performance at $\rho=0.85$, $n_3=200$, $k=3$,
$w_3=3$, $B=500$ replications. ARI and $F_1$ (higher is better) are shown
for both methods; Specificity, Precision, Balanced Accuracy, and AUC
(all higher is better) are shown for FSC-NS only (FSC-SRS
Specificity is near zero for small $\varepsilon$ and is
uninformative). Bold marks the highest FSC-NS ARI in each $\varepsilon$ block;
\textit{italics} mark FSC-SRS ARI entries at or below $0.05$, indicating
near-degenerate all-rare prediction.}
\label{tab:rare-classif}
\setlength{\tabcolsep}{4pt}
{%
\begin{tabular}{@{}ccc cc c cccc@{}}
\toprule
& & &
\multicolumn{2}{c}{ARI $\uparrow$} &
$F_1$ $\uparrow$ &
\multicolumn{4}{c}{FSC-NS only} \\
\cmidrule(lr){4-5}\cmidrule(lr){6-6}\cmidrule(lr){7-10}
$\varepsilon$ & $\delta$ & $\tau$ &
NS & SRS & NS &
Spec.$\uparrow$ & Prec.$\uparrow$ & Bal.Acc.$\uparrow$ & AUC$\uparrow$ \\
\midrule
0.02 & 3 & 1.0 & 0.645 & \textit{0.008} & 0.687 & 0.993 & 0.846 & 0.796 & 0.972 \\
     & 3 & 1.5 & 0.630 & \textit{0.023} & 0.671 & 0.994 & 0.873 & 0.779 & 0.947 \\
     & 3 & 2.0 & 0.657 & 0.094 & 0.687 & 0.995 & 0.891 & 0.790 & 0.936 \\
     & 4 & 1.0 & 0.866 & 0.208 & 0.885 & 0.996 & 0.932 & 0.924 & 0.995 \\
     & 4 & 1.5 & 0.809 & 0.263 & 0.832 & 0.996 & 0.934 & 0.882 & 0.984 \\
     & 4 & 2.0 & 0.771 & 0.413 & 0.796 & 0.996 & 0.933 & 0.856 & 0.966 \\
     & 5 & 1.0 & \textbf{0.957} & 0.718 & 0.963 & 0.998 & 0.970 & 0.979 & 1.000 \\
     & 5 & 1.5 & 0.908 & 0.671 & 0.922 & 0.998 & 0.964 & 0.944 & 0.995 \\
     & 5 & 2.0 & 0.878 & 0.705 & 0.894 & 0.997 & 0.955 & 0.924 & 0.988 \\
\midrule
0.05 & 3 & 1.0 & 0.704 & \textit{$-$0.014} & 0.782 & 0.982 & 0.869 & 0.852 & 0.972 \\
     & 3 & 1.5 & 0.680 &  \textit{0.046}  & 0.758 & 0.987 & 0.890 & 0.828 & 0.950 \\
     & 3 & 2.0 & 0.682 &  0.204  & 0.756 & 0.987 & 0.890 & 0.828 & 0.935 \\
     & 4 & 1.0 & 0.888 &  0.283  & 0.924 & 0.991 & 0.944 & 0.950 & 0.995 \\
     & 4 & 1.5 & 0.829 &  0.417  & 0.880 & 0.992 & 0.944 & 0.911 & 0.984 \\
     & 4 & 2.0 & 0.788 &  0.527  & 0.847 & 0.991 & 0.934 & 0.886 & 0.969 \\
     & 5 & 1.0 & \textbf{0.963} & 0.790 & 0.976 & 0.996 & 0.979 & 0.985 & 1.000 \\
     & 5 & 1.5 & 0.916 &  0.711  & 0.943 & 0.995 & 0.969 & 0.958 & 0.996 \\
     & 5 & 2.0 & 0.875 &  0.732  & 0.914 & 0.994 & 0.961 & 0.934 & 0.987 \\
\midrule
0.10 & 3 & 1.0 & 0.699 & \textit{$-$0.028} & 0.844 & 0.964 & 0.888 & 0.886 & 0.971 \\
     & 3 & 1.5 & 0.665 &  \textit{0.038}  & 0.813 & 0.972 & 0.902 & 0.859 & 0.951 \\
     & 3 & 2.0 & 0.659 &  0.261  & 0.804 & 0.972 & 0.899 & 0.854 & 0.938 \\
     & 4 & 1.0 & 0.884 &  0.354  & 0.946 & 0.984 & 0.956 & 0.961 & 0.995 \\
     & 4 & 1.5 & 0.808 &  0.400  & 0.906 & 0.981 & 0.945 & 0.927 & 0.984 \\
     & 4 & 2.0 & 0.770 &  0.500  & 0.882 & 0.979 & 0.937 & 0.908 & 0.971 \\
     & 5 & 1.0 & \textbf{0.963} & 0.834 & 0.983 & 0.995 & 0.985 & 0.988 & 0.999 \\
     & 5 & 1.5 & 0.912 &  0.691  & 0.959 & 0.990 & 0.973 & 0.969 & 0.996 \\
     & 5 & 2.0 & 0.865 &  0.702  & 0.936 & 0.988 & 0.966 & 0.948 & 0.988 \\
\bottomrule
\end{tabular}}
\end{table}

\begin{table}[tbp]
\centering
\caption{Effect of set size $k$ at $\rho=0.85$, $\varepsilon=0.05$,
$\delta=4$, $\tau=1.5$, $n_3=200$, $w_3=3$, $B=500$ replications.
ARI and $F_1$ are higher-is-better. FSC-SRS sensitivity near 1 at
$k\ge5$ is degenerate (see text).}
\label{tab:rare-k-effect}
\setlength{\tabcolsep}{5pt}
\begin{tabular}{@{}c cccc cccc@{}}
\toprule
& \multicolumn{4}{c}{{FSC-NS}}
& \multicolumn{4}{c}{{FSC-SRS}} \\
\cmidrule(lr){2-5}\cmidrule(lr){6-9}
$k$ & ARI & Sens. & $F_1$ & Iter.
    & ARI & Sens. & $F_1$ & Iter. \\
\midrule
2 & 0.834 & 0.820 & 0.869 & 18.4 & 0.745 & 0.916 & 0.809 &  48.5 \\
3 & 0.829 & 0.829 & 0.880 & 19.5 & 0.417 & 0.972 & 0.629 &  59.5 \\
5 & 0.807 & 0.846 & 0.890 & 20.5 & $-$0.026 & 0.999 & 0.397 & 122.0 \\
8 & 0.789 & 0.870 & 0.908 & 19.9 & 0.000 & 1.000 & 0.474 & 207.9 \\
\bottomrule
\end{tabular}
\end{table}

\begin{table}[tbp]
\centering
\caption{Effect of ranking quality $\rho$ on classification at $n_3=200$,
$k=3$, $w_3=3$, averaged over all 27 $(\varepsilon,\delta,\tau)$ combinations.
All metrics are higher-is-better. FSC-NS is stable across all three
$\rho$ levels; FSC-SRS improves as ranking degrades only because the NS
enrichment is attenuated.}
\label{tab:rare-imperfect-classif}
\setlength{\tabcolsep}{4.5pt}
\begin{tabular}{@{}cc cccccc@{}}
\toprule
$\rho$ & Method & ARI & Sens. & Spec. & Prec. & $F_1$ & Bal.Acc. \\
\midrule
1.00 & FSC-NS & 0.796 & 0.808 & 0.988 & 0.926 & 0.855 & 0.898 \\
     & FSC-SRS        & 0.256 & 0.983 & 0.438 & 0.378 & 0.479 & 0.711 \\
\midrule
0.85 & FSC-NS & 0.803 & 0.811 & 0.989 & 0.931 & 0.859 & 0.900 \\
     & FSC-SRS        & 0.391 & 0.969 & 0.637 & 0.486 & 0.585 & 0.803 \\
\midrule
0.60 & FSC-NS & 0.816 & 0.816 & 0.991 & 0.938 & 0.865 & 0.904 \\
     & FSC-SRS        & 0.647 & 0.935 & 0.896 & 0.688 & 0.760 & 0.915 \\
\bottomrule
\end{tabular}
\end{table}

\begin{figure}[tbp]
\centering
\caption{Average ARI versus $w_3$ at $k=3$, $n_3=200$, averaged over all
$(\varepsilon,\delta,\tau)$ combinations. Left: FSC-NS; right:
FSC-SRS. Curves correspond to ranking levels $\rho\in\{1,0.85,0.60\}$.
ARI$\in[-1,1]$; higher is better. The three curves in the left panel
nearly coincide, confirming that FSC-NS is robust to ranking
quality. FSC-SRS deteriorates with $w_3$ most severely under $\rho=1$.}
\includegraphics[width=\textwidth]{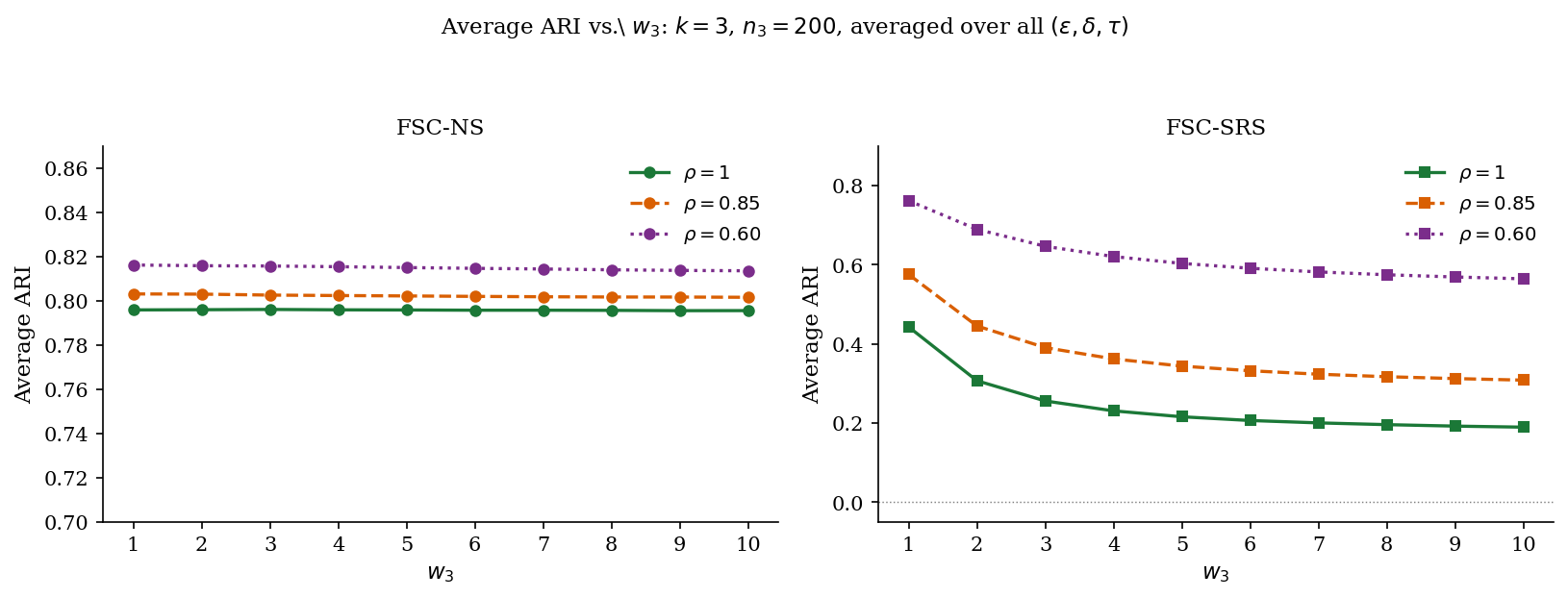}
\label{fig:rare-imperfect-ari-w3}
\end{figure}

\begin{figure}[tbp]
\centering
\caption{Average rare-class $F_1$ score versus $w_3$ under the same
averaging scheme as Figure~\ref{fig:rare-imperfect-ari-w3}.
$F_1\in[0,1]$; higher is better. FSC-NS is flat and nearly
$\rho$-invariant; FSC-SRS degrades monotonically with $w_3$, more
severely for accurate ranking.}
\includegraphics[width=\textwidth]{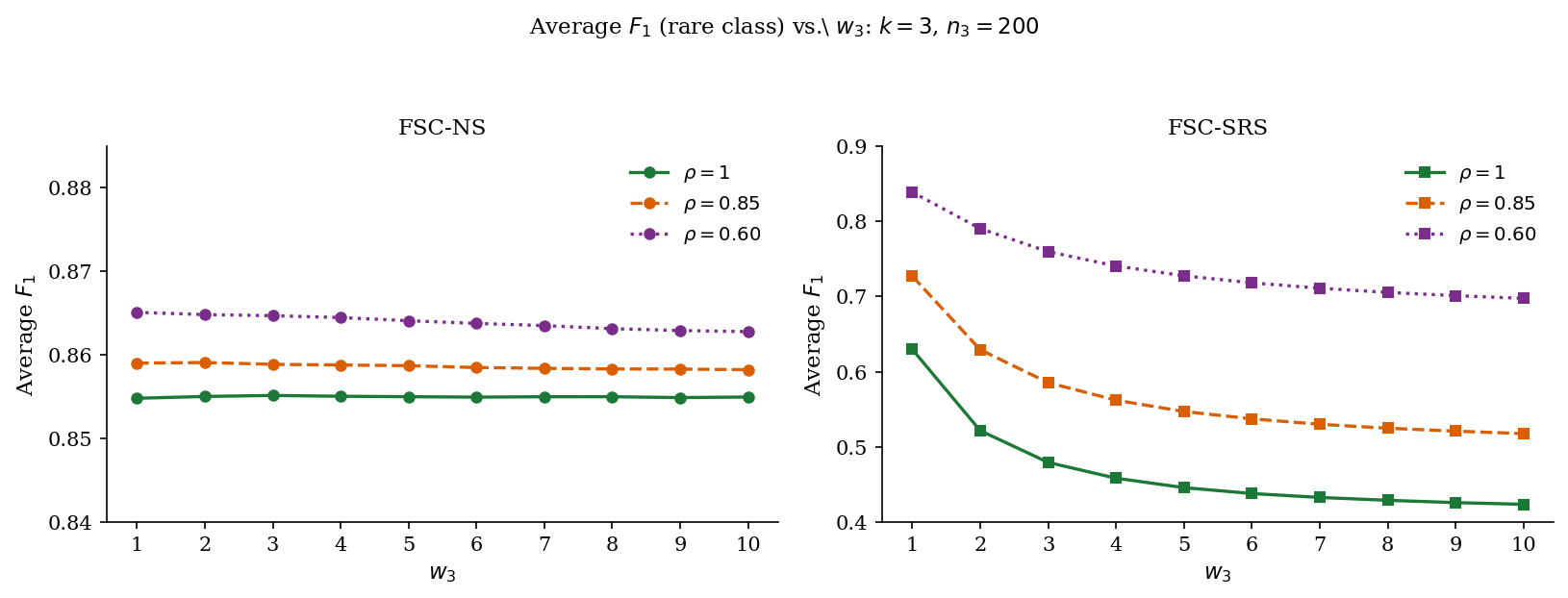}
\label{fig:rare-imperfect-f1-w3}
\end{figure}

\medskip
\noindent\textbf{Effect of ranking quality:}
Table~\ref{tab:rare-imperfect-classif} compares the three ranking levels,
averaging over all 27 $(\varepsilon,\delta,\tau)$ combinations at $n_3=200$,
$k=3$, $w_3=3$. FSC-NS is remarkably stable: its mean ARI changes
from $0.796$ ($\rho=1$) to $0.803$ ($\rho=0.85$) to $0.816$ ($\rho=0.60$),
and its mean $F_1$ from $0.855$ to $0.859$ to $0.865$. The method tolerates
imperfect ranking because the working NS likelihood is misspecified only in
the degree to which the selected unit truly is the set maximum; when the
ranking score is noisy the nominated unit is still enriched for the rare class,
just less strongly so.
FSC-SRS exhibits the reverse monotone pattern. Its mean ARI rises from
$0.256$ to $0.391$ to $0.647$ as $\rho$ decreases. This apparent improvement
does not reflect better classification; it reflects attenuation of the NS
enrichment. As ranking becomes noisier, the nominated data look more like an
ordinary SRS sample, so the SRS likelihood is less misspecified. Stated simply,
{the better the ranker, the greater the benefit of correctly modeling
the nomination mechanism and the more severe the cost of ignoring it}.
It is important to note that both methods are fitted
throughout under the working assumption of perfect ranking.  Under imperfect
ranking, therefore, FSC-NS is also technically misspecified, since the
selected unit is not guaranteed to be the set maximum.  The results confirm
that FSC-NS tolerates this working misspecification gracefully as its
performance is essentially unchanged across $\rho$ levels. On the other hand,  FSC-SRS
is doubly misspecified as not only it wrongly treats enriched data as SRS draws but also that  error is not eliminated as ranking degrades.  The
robustness of FSC-NS to ranking noise is a practically important finding,
since perfect ranking is rarely achievable in field applications.
Figures~\ref{fig:rare-imperfect-ari-w3}--\ref{fig:rare-imperfect-epsrmse-w3}
display the three $\rho$ levels as $w_3$ varies. For FSC-NS the three
curves are essentially coincident and flat across $w_3$, confirming that
increasing the unlabeled weight consistently helps regardless of ranking
quality. For FSC-SRS the deterioration with $w_3$ is monotone and steepest
under $\rho=1$, because more weight on perfectly ranked but misspecified data
amplifies the likelihood misspecification most severely.

\begin{figure}[tbp]
\centering
\caption{Average RMSE of $\hat\varepsilon$ versus $w_3$ at $k=3$,
$n_3=200$. Lower is better. FSC-NS stays near zero across
$w_3$ and $\rho$. FSC-SRS grows with $w_3$, most severely under $\rho=1$.}
\includegraphics[width=\textwidth]{eps_rmse_vs_w3.png}
\label{fig:rare-imperfect-epsrmse-w3}
\end{figure}

\medskip
\noindent\textbf{Parameter estimation:}
Table~\ref{tab:rare-params-compact} reports bias and RMSE for
$(\hat\varepsilon,\hat\delta,\hat\tau)$ at $\rho=0.85$, $n_3=200$, $k=3$,
$w_3=3$, for nine representative $(\varepsilon,\delta)$ pairs at $\tau=1.5$.
FSC-NS estimates $\varepsilon$ with near-zero bias throughout (RMSE
$0.006$--$0.063$); the correct denominator $n_3k$ in the M-step removes the
selection bias inherent in the NS mechanism. Estimation of $\delta$ and
$\tau$ becomes harder as $\tau$ increases (wider rare component) and easier as
$\delta$ increases (stronger signal), as expected.
FSC-SRS is severely miscalibrated: bias in $\hat\varepsilon$ ranges from
$0.17$ to $0.88$, so the estimated contamination fraction can exceed the true
value by a factor of four to forty. This propagates directly into the posterior
probabilities and explains the degenerate classification of
Table~\ref{tab:rare-classif}.
Table~\ref{tab:rare-imperfect-estimation} shows the effect of $\rho$ on
estimation, averaged over all 27 configurations. For FSC-NS, the bias
in $\hat\varepsilon$ stays below $0.01$ at every ranking level, and RMSE grows
only from $0.014$ to $0.054$ as $\rho$ drops from $1$ to $0.60$; mean EM
iterations fall from $25.1$ to $17.8$, since noisier ranking produces less
informative data that converge faster. For FSC-SRS, bias and RMSE in
$\hat\varepsilon$ decrease monotonically with $\rho$ ($0.597\to0.437\to0.217$
and $0.629\to0.466\to0.237$) for the same attenuation reason, but the method
remains substantially worse than FSC-NS at every level.

\begin{table}[tbp]
\centering
\caption{Parameter bias and RMSE (lower is better) at $\rho=0.85$,
$n_3=200$, $k=3$, $w_3=3$. Nine representative $(\varepsilon,\delta)$
configurations at $\tau=1.5$; the full 27-configuration pattern is the same.}
\label{tab:rare-params-compact}
\setlength{\tabcolsep}{4pt}
{%
\begin{tabular}{@{}ccc cc cc cc@{}}
\toprule
& & &
\multicolumn{2}{c}{$\hat\varepsilon$} &
\multicolumn{2}{c}{$\hat\delta$} &
\multicolumn{2}{c}{$\hat\tau$} \\
\cmidrule(lr){4-5}\cmidrule(lr){6-7}\cmidrule(lr){8-9}
$\varepsilon$ & $\delta$ & $\tau$ &
Bias & RMSE & Bias & RMSE & Bias & RMSE \\
\midrule
\multicolumn{9}{l}{{FSC-NS}} \\
\midrule
0.02 & 3 & 1.5 & $\phantom{-}$0.003 & 0.037 & $\phantom{-}$0.045 & 0.934 & $-$0.086 & 0.473 \\
0.02 & 4 & 1.5 & $\phantom{-}$0.001 & 0.020 & $\phantom{-}$0.065 & 0.724 & $-$0.102 & 0.404 \\
0.02 & 5 & 1.5 & $-$0.001 & 0.006 & $\phantom{-}$0.033 & 0.371 & $-$0.074 & 0.300 \\
0.05 & 3 & 1.5 & $\phantom{-}$0.003 & 0.049 & $\phantom{-}$0.095 & 0.911 & $-$0.094 & 0.450 \\
0.05 & 4 & 1.5 & $-$0.002 & 0.021 & $\phantom{-}$0.117 & 0.517 & $-$0.112 & 0.317 \\
0.05 & 5 & 1.5 & $-$0.001 & 0.010 & $\phantom{-}$0.021 & 0.321 & $-$0.058 & 0.239 \\
0.10 & 3 & 1.5 & $-$0.001 & 0.063 & $\phantom{-}$0.117 & 0.775 & $-$0.094 & 0.389 \\
0.10 & 4 & 1.5 & $-$0.004 & 0.016 & $\phantom{-}$0.092 & 0.279 & $-$0.076 & 0.203 \\
0.10 & 5 & 1.5 & $-$0.000 & 0.014 & $\phantom{-}$0.011 & 0.233 & $-$0.038 & 0.181 \\
\midrule
\multicolumn{9}{l}{{FSC-SRS}} \\
\midrule
0.02 & 3 & 1.5 & $\phantom{-}$0.876 & 0.901 & $-$1.927 & 2.003 & $-$0.370 & 0.396 \\
0.02 & 4 & 1.5 & $\phantom{-}$0.463 & 0.547 & $-$2.088 & 2.280 & $\phantom{-}$0.021 & 0.302 \\
0.02 & 5 & 1.5 & $\phantom{-}$0.166 & 0.218 & $-$1.540 & 1.967 & $\phantom{-}$0.428 & 0.613 \\
0.05 & 3 & 1.5 & $\phantom{-}$0.659 & 0.688 & $-$1.524 & 1.558 & $-$0.115 & 0.185 \\
0.05 & 4 & 1.5 & $\phantom{-}$0.355 & 0.370 & $-$1.590 & 1.650 & $\phantom{-}$0.307 & 0.345 \\
0.05 & 5 & 1.5 & $\phantom{-}$0.227 & 0.238 & $-$1.444 & 1.597 & $\phantom{-}$0.624 & 0.682 \\
0.10 & 3 & 1.5 & $\phantom{-}$0.580 & 0.591 & $-$1.098 & 1.123 & $\phantom{-}$0.055 & 0.117 \\
0.10 & 4 & 1.5 & $\phantom{-}$0.418 & 0.425 & $-$1.201 & 1.233 & $\phantom{-}$0.408 & 0.417 \\
0.10 & 5 & 1.5 & $\phantom{-}$0.316 & 0.322 & $-$1.055 & 1.154 & $\phantom{-}$0.633 & 0.664 \\
\bottomrule
\end{tabular}}
\end{table}

\begin{table}[tbp]
\centering
\caption{Effect of ranking quality on parameter estimation at $n_3=200$,
$k=3$, $w_3=3$, averaged over all 27 $(\varepsilon,\delta,\tau)$ combinations.
Bias and RMSE are lower-is-better; Iter.\ is mean EM iterations.}
\label{tab:rare-imperfect-estimation}
\setlength{\tabcolsep}{4pt}
\begin{tabular}{@{}cc cccccc c@{}}
\toprule
$\rho$ & Method &
Bias$(\hat\varepsilon)$ & RMSE$(\hat\varepsilon)$ &
Bias$(\hat\delta)$ & RMSE$(\hat\delta)$ &
Bias$(\hat\tau)$ & RMSE$(\hat\tau)$ & Iter. \\
\midrule
1.00 & FSC-NS & $\phantom{-}$0.001 & 0.014 & $\phantom{-}$0.013 & 0.391 & $-$0.044 & 0.273 &  25.1 \\
     & FSC-SRS        & $\phantom{-}$0.597 & 0.629 & $-$1.694 & 1.838 & $\phantom{-}$0.111 & 0.446 & 137.5 \\
\midrule
0.85 & FSC-NS & $\phantom{-}$0.004 & 0.030 & $-$0.036 & 0.732 & $-$0.040 & 0.383 &  25.5 \\
     & FSC-SRS        & $\phantom{-}$0.437 & 0.466 & $-$1.368 & 1.538 & $\phantom{-}$0.193 & 0.436 & 114.2 \\
\midrule
0.60 & FSC-NS & $\phantom{-}$0.008 & 0.054 & $-$0.061 & 1.142 & $-$0.066 & 0.498 &  17.8 \\
     & FSC-SRS        & $\phantom{-}$0.217 & 0.237 & $-$0.658 & 0.962 & $\phantom{-}$0.193 & 0.381 &  57.6 \\
\bottomrule
\end{tabular}
\end{table}

%\begin{figure}[tbp]
%\centering
%\caption{Average ARI versus $w_3$ at $\rho=0.85$, $\varepsilon=0.05$,
%$\delta=4$, $\tau=1.5$, $n_3=200$. Panels show set sizes $k\in\{2,3,5,8\}$.
%ARI$\in[-1,1]$; higher is better. FSC-NS (solid, circles) stays
%near $0.83$ across all $k$ and weights. FSC-SRS (dashed, squares)
%deteriorates sharply for $k\ge3$ as $w_3$ grows, with ARI near zero
%at $k=5$ and $k=8$ even for moderate weights.}
%\includegraphics[width=\textwidth]{ARI_eps05_d4_t15_n200_rho085.png}
%\label{fig:rare-ari}
%\end{figure}
%
%\begin{figure}[tbp]
%\centering
%\caption{Sensitivity for the rare class versus $w_3$ at $\rho=0.85$
%under the same scenario as Figure~\ref{fig:rare-ari}. Sensitivity$\in[0,1]$;
%higher is better. FSC-SRS near-perfect sensitivity at $k\ge3$ is degenerate:
%it arises from labeling almost everything as rare, as confirmed by the
%near-zero ARI and collapsed specificity. FSC-NS maintains calibrated
%sensitivity near $0.83$ with specificity above $0.99$.}
%\includegraphics[width=\textwidth]{Sens_eps05_d4_t15_n200_rho085.png}
%\label{fig:rare-sens}
%\end{figure}

In summary, FSC-NS delivers well-calibrated estimates and reliable
rare-class recovery across all $(\varepsilon,\delta,\tau,k,w_3,\rho)$
combinations studied. Under mild ranking error ($\rho=0.85$) its performance
is virtually unchanged from perfect ranking, while FSC-SRS remains
structurally misspecified and its classification degrades to near-degenerate
all-rare prediction whenever nomination enrichment is substantial. The better
the ranker, the greater the cost of ignoring the nomination mechanism.
On the practical question of how to choose $w_3$, the simulation results
suggest a useful interim recommendation. Under FSC-NS, performance is stable
across a wide range of moderate $w_3$ values: in Figure~\ref{fig:rare-ari-sens},
ARI varies little for $w_3\in\{1,\ldots,6\}$ at every $k$. Under FSC-SRS,
by contrast, large $w_3$ amplifies the misspecification severely. In the
absence of a principled selection procedure for the NS setting, we recommend choosing $w_3$ by cross-validated
predictive log-likelihood or ARI on a small labeled holdout when one is
available, or defaulting to moderate values such as $w_3\in\{2,4\}$ when no
holdout is available. Values of $w_3\geq8$ should be avoided unless the
unlabeled sample is known to be representative.

%=========================================================================

\section{Real-Data Illustration}
\label{sec:realdata}
%=========================================================================

We illustrate the proposed methodology using the Wisconsin Diagnostic
Breast Cancer (WDBC) dataset \citep{wdbc_dataset,mangasarian1995}, a
widely studied binary classification benchmark that contains
morphological measurements computed from digitized fine-needle aspirate
(FNA) images of breast masses. Each record corresponds to a distinct
patient biopsy and carries a confirmed diagnosis of either malignant
(M) or benign (B), yielding a total of $569$ observations with $212$
malignant and $357$ benign cases.  The WDBC dataset has served as a
standard testbed for classification methodology, including mixture
model approaches \citep{bouveyron2019}.

This section is best understood as a {data-based illustration
under an imposed NS protocol} rather than an analysis of a
prospectively collected nominated sample.  The NS structure is
constructed post-hoc from an existing dataset in order to isolate and
demonstrate the design-correction principle in a controlled way;
the aim is not to optimize breast-cancer classification per se, but
to show that the correct likelihood formulation matters when nominated
data are present.  Our analysis is further restricted to the
univariate setting consistent with the theoretical development of
this paper, and the univariate reduction is deliberate: it allows
each component's contribution to be traced directly through the
likelihood expressions derived in Section~\ref{sec:background}.

We take the measured variable to be $Y = \log(\text{area\_worst})$, where
$\text{area\_worst}$ is the largest observed area among the three
cell-nucleus measurements recorded per observation; this transformation
renders the distributions of the two diagnostic groups visually closer
to Gaussian.  The ranking variable used to construct the NS sets is
$\text{radius\_worst}$, the largest recorded nuclear radius, which is
strongly correlated with area and provides a readily available
auxiliary ordering criterion \citep{mangasarian1995}.

In each replicated experiment, we draw $n_1=n_2=20$ labeled
observations from the malignant and benign classes, respectively, and
an unlabeled sample of size $n_3=80$ from the combined pool, all
without replacement. Within each group of $k$ observations, the unit with the
largest value of $\text{radius\_worst}$ is designated as the nominated
(maximum-ranked) observation, and only that unit's $Y$ value is
retained for the labeled or unlabeled sample.  Set sizes
$k\in\{2,3,4\}$ and unlabeled-data weights
$w_3\in\{0,2,4,6,8,10\}$ (with $w_1=w_2=1$ throughout) are
considered.  All results are averaged over $500$ independent
replications of this sampling process.

\medskip
\noindent\textbf{Competing methods:}
The empirical study contrasts  two  classification procedures.
{FSC-NS} is the proposed method, which uses the proper
latent pair $(Z_r,V_r)$ as derived in Section~\ref{sec:background}.
{FSC-SRS} is the ordinary SRS-based FSC procedure of
\citet{vrbik}, which treats the retained nominated values as if they
were simple random draws from a Gaussian mixture.  Both methods
are initialized identically and converge under the same stopping
criterion, ensuring that performance differences reflect the likelihood
specification rather than algorithmic choices.

For each configuration we record the average adjusted Rand index, misclassification error,
sensitivity, specificity, precision, $F_1$ score, balanced accuracy,
AUC, log-loss, the estimated mixing
proportion $\hat{\pi}$, and the mean EM iteration count over the $500$
replications.

\medskip
\noindent\textbf{Performance of the proposed method:}
Table~\ref{tab:wdbc_correct} presents the full profile of the proposed
FSC-NS procedure across all $(k, w_3)$ configurations.
Several coherent patterns emerge.
\begin{itemize}
\item First, moderate fractional use of the unlabeled sample consistently
improves classification relative to the purely supervised baseline
($w_3=0$).  The benefit is most pronounced as $k$ increases: at
$k=4$, for instance, the average ARI rises from $0.5874$ at $w_3=0$
to $0.6431$ at $w_3=4$, a gain of nearly five ARI points, while the
misclassification error falls from $0.1038$ to $0.0924$.  This pattern
is consistent with the theoretical efficiency advantage of NS over
SRS: larger set sizes make the nominated maximum more informative
about the underlying component structure, and the correct likelihood
exploits that information through the $\tilde v_r$ contribution.

\item Second, the optimal configuration across all $(k, w_3)$ combinations
is $k=4$, $w_3=4$, which yields ARI $=0.6431$, error $=0.0924$,
sensitivity $=0.9142$, specificity $=0.8846$, balanced accuracy
$=0.8994$, and AUC $=0.9676$.  The high AUC is particularly notable:
an area under the ROC curve of $0.9676$ indicates that the
proposed procedure can reliably rank malignant observations above
benign ones, a property of direct clinical relevance in diagnostic
screening applications \citep{hanley1982,pepe2003}.

\item Third, excessive weighting of the unlabeled observations ($w_3=10$)
severely degrades performance for all three values of $k$.  The
procedure becomes effectively unsupervised in this regime, and the
average ARI collapses to $0.4667$, $0.5060$, and $0.5461$ for
$k=2,3,4$, respectively, with misclassification errors exceeding
$0.83$.  This collapse mirrors the behavior documented in the SRS
simulations and in the broader FSC literature \citep{cozman(2003),vrbik},
and confirms that the nominated-data setting inherits the same
fundamental trade-off between the stabilizing influence of the
labeled sample and the global information contributed by the unlabeled
observations.

\item Fourth, the computational cost of the algorithm grows with $w_3$ in a
predictable manner.  At the best-performing configuration
$(k=4, w_3=4)$, the mean iteration count is $71.3$; this rises to
$182.0$ at $w_3=8$ and then decreases to $172.2$ at $w_3=10$, where
rapid convergence to a degenerate solution accounts for the reduction.
The observed iteration counts are consistent with the EM convergence
theory of \citet{Dempster(1977)}, and the algorithm remains
computationally feasible throughout the weight range considered.
\end{itemize} 
\begin{table}[tb!]
\centering
\caption{Average classification and estimation summaries for the
proposed FSC-NS procedure on the WDBC data, based on $500$
replications. The NS samples are formed using
$\log(\text{area\_worst})$ as the measured variable and
$\text{radius\_worst}$ as the ranking variable, with $n_1=n_2=20$
labeled observations and $n_3=80$ unlabeled observations per
replicate.}
{
\begin{tabular}{cccccccccc}
\toprule
$k$ & $w_3$ & ARI & Error & Sensitivity & Specificity & Bal.\ Acc. & AUC & $\hat{\pi}$ & Iter. \\
\midrule
2 & 0  & 0.5759 & 0.1201 & 0.9115 & 0.8363 & 0.8739 & 0.9639 & 0.5216 &   2.0 \\
2 & 2  & 0.6043 & 0.1112 & 0.8540 & 0.9307 & 0.8924 & 0.9640 & 0.3424 &  30.2 \\
2 & 4  & 0.5973 & 0.1137 & 0.8436 & 0.9382 & 0.8909 & 0.9638 & 0.3372 &  57.1 \\
2 & 6  & 0.5850 & 0.1181 & 0.8359 & 0.9397 & 0.8878 & 0.9636 & 0.3361 &  94.6 \\
2 & 8  & 0.5445 & 0.1330 & 0.8236 & 0.9236 & 0.8736 & 0.9626 & 0.3425 & 153.2 \\
2 & 10 & 0.4667 & 0.8368 & 0.2592 & 0.0386 & 0.1489 & 0.0373 & 0.7293 & 125.8 \\
\midrule
3 & 0  & 0.5830 & 0.1141 & 0.9356 & 0.7771 & 0.8564 & 0.9651 & 0.5121 &   2.0 \\
3 & 2  & 0.6180 & 0.1048 & 0.8937 & 0.8963 & 0.8950 & 0.9651 & 0.3321 &  35.4 \\
3 & 4  & 0.6192 & 0.1048 & 0.8868 & 0.9118 & 0.8993 & 0.9651 & 0.3264 &  64.8 \\
3 & 6  & 0.6062 & 0.1094 & 0.8752 & 0.9234 & 0.8993 & 0.9651 & 0.3174 & 106.2 \\
3 & 8  & 0.5770 & 0.1194 & 0.8578 & 0.9316 & 0.8947 & 0.9650 & 0.3088 & 175.9 \\
3 & 10 & 0.5060 & 0.8536 & 0.1899 & 0.0507 & 0.1203 & 0.0357 & 0.7252 & 149.8 \\
\midrule
4 & 0  & 0.5874 & 0.1038 & 0.9520 & 0.7230 & 0.8375 & 0.9676 & 0.5041 &   2.0 \\
4 & 2  & 0.6396 & 0.0929 & 0.9180 & 0.8704 & 0.8942 & 0.9676 & 0.3177 &  40.1 \\
4 & 4  & 0.6431 & 0.0924 & 0.9142 & 0.8846 & 0.8994 & 0.9676 & 0.3126 &  71.3 \\
4 & 6  & 0.6386 & 0.0944 & 0.9063 & 0.9006 & 0.9035 & 0.9676 & 0.3036 & 113.2 \\
4 & 8  & 0.6177 & 0.1020 & 0.8907 & 0.9189 & 0.9048 & 0.9676 & 0.2895 & 182.0 \\
4 & 10 & 0.5461 & 0.8717 & 0.1475 & 0.0675 & 0.1075 & 0.0333 & 0.7341 & 172.2 \\
\bottomrule
\end{tabular}}
\label{tab:wdbc_correct}
\end{table}

\begin{table}[tb!]
\centering
\caption{Comparison of FSC-NS and FSC-SRS on
the WDBC data at the best-performing configuration ($k=4$, $w_3=4$).
All entries are averaged over $500$ replications.}
\resizebox{0.97\textwidth}{!}{
\begin{tabular}{lccccccccc}
\toprule
Method & ARI & Error & Sensitivity & Specificity & Precision & $F_1$ & Bal.\ Acc. & AUC & Iter. \\
\midrule
FSC-NS & 0.6431 & 0.0924 & 0.9142 & 0.8846 & 0.9635 & 0.9371 & 0.8994 & 0.9676 & 71.3 \\
FSC-SRS        & 0.5629 & 0.1210 & 0.8546 & 0.9531 & 0.9847 & 0.9135 & 0.9039 & 0.9676 & 37.0 \\
\bottomrule
\end{tabular}
}
\label{tab:wdbc_compare}
\end{table}

 For the proposed method, the estimated mixing
proportion decreases steadily from approximately $0.52$, $0.51$, and
$0.50$ under $w_3=0$ to approximately $0.34$, $0.33$, and $0.31$ at
the best-performing partially supervised configurations for
$k=2,3,4$, respectively.  This systematic downward shift arises
because the NS mechanism over-represents the higher-valued
component in the unlabeled sample. By retaining only the maximum
within each set, the nominated observations are disproportionately
drawn from the malignant group, which has both a higher mean and a
larger variance in $\log(\text{area\_worst})$.  The proposed method
correctly accounts for this bias through the $\tilde v_r$ cdf term,
so the resulting estimate of $\pi$ reflects the effective
representation of each class in the nominated rather than the natural
sample.  The importance of correcting for such design-induced
representation bias in partially labeled data has been noted in the
semi-supervised classification literature
\citep{dean2006,castelli1996}, and the present results show that the
NS correction achieves this automatically through the proper
likelihood.
Together, these findings establish that a correct likelihood formulation
for the nomination-sampling design is not merely a theoretical
refinement but a practical necessity: it determines whether partially
labeled nominated data can be exploited at all, rather than
simply how efficiently.

%=========================================================================

\section{Discussion and Conclusion}
\label{sec:conclusion}
%=========================================================================

The central message of this paper is that the sampling design must
enter the FSC likelihood explicitly once the partially labeled data
have been collected under nomination sampling.  The standard
single-indicator EM construction is fundamentally incompatible with
the NS observation mechanismThe
proposed latent pair $(Z_i,V_i)$ restores this connection and
thereby yields a valid weighted-likelihood FSC procedure for
nominated data, one that reduces exactly to ordinary FSC-SRS when
$k=1$.

Two structural features of the resulting algorithm deserve
emphasis.  The E-step decomposes naturally into a density-based
term $\tilde z_r$ for the observed maximum and a cdf-based term
$\tilde v_r$ for the unmeasured remainder of the set.  This
decomposition is the statistical signature of the nomination
mechanism. It encodes the two distinct pieces of latent information
carried by a single nominated observation, whereas the naive
single-indicator construction discards the second.  The M-step
update for the mixing proportion reflects this as well, using a
denominator of $n_3 k$ rather than $n_3$ to acknowledge that each
retained observation represents $k$ latent draws.  These are not
incidental adjustments but the structural changes that align the
EM iterations with the true likelihood.

The simulation study demonstrates that these differences matter
substantially in practice.  The misspecified EM produces biases
of order one to two units across the weight range studied, and its
bias grows with $w_3$ as more weight is placed on the misspecified
unlabeled likelihood, confirming that the failure is a property of
the objective function rather than the data.  When $w_3=0$ the
unlabeled contribution is absent and both methods coincide, so the
misspecification is harmless at that boundary; it becomes progressively
more damaging as $w_3$ increases.  The
proposed method, by contrast, shows calibrated estimation accuracy
across the moderate weight range and achieves ARI values that
exceed those based on the standard  SRS framework.

 The present
development is confined to two-component mixtures and a fixed,
common set size $k$, both chosen to make the latent-variable
structure as transparent as possible.  Extensions to $G>2$
components are algebraically straightforward but require more careful treatment of label
constraints.
Variable set sizes, which arise when some sets are incomplete due
to measurement constraints, can be accommodated by conditioning
the cdf factor on the realized set size; again, the structural
argument is clear but the details warrant a separate treatment.
Formal asymptotic theory for the weighted FSC-NS estimators,
including rates of convergence and an analogue of the
information-theoretic analysis of \citet{castelli1996} for the
nominated-data setting, represents another important open
direction.  On the applied side, adapting the weight-selection
procedures of \citet{gallaugher2019} to the NS context would make
the methodology directly operational, removing the need to specify
$w_3$ by cross-validation or expert judgment.

%%--------------------------------------------------------------
%%  SECTION 6: DISCUSSION
%%--------------------------------------------------------------
\section*{Acknowledgment}
Mohammad Jafari Jozani gratefully acknowledges partial support from NSERC Canada.

%%--------------------------------------------------------------
%%  BIBLIOGRAPHY
%%--------------------------------------------------------------

%\bibliographystyle{apalike}
%\bibliography{references_NS_v2_clean}

\end{document}